\documentclass[useAMS,usenatbib]{mnras}

\usepackage{rotating}
\usepackage{lscape}
\usepackage{graphicx}
\usepackage{amssymb}
\usepackage{amsmath}
\usepackage{epstopdf}
\usepackage{color,colortbl}
\usepackage{soul}
\usepackage{multirow}
\usepackage{longtable}
\usepackage{textcomp}
\usepackage[caption=false]{subfig}
\usepackage{float}
\usepackage{appendix}
\usepackage{listings}

 \def\aap{Astron. Astrophys.}
 \def\apjs{Astrophys. J., Suppl. Ser.}

\definecolor{dkgreen}{rgb}{0,0.6,0}
\definecolor{gray}{rgb}{0.5,0.5,0.5}
\definecolor{mauve}{rgb}{0.58,0,0.82}

\pdfoutput=1

\usepackage{hyperref}
\title[HELP:XID+]
  {HELP: XID+, The Probabilistic De-blender for \emph{Herschel} SPIRE maps}\author[P.D. Hurley et al.]{P.D.~Hurley,$^1$\thanks{Email: p.d.hurley@sussex.ac.uk} S.~Oliver,$^1$ M.~Betancourt,$^2$ C.~Clarke,$^1$ W.I.~Cowley,$^3$\newauthor S.~Duivenvoorden,$^1$ 
 D.~Farrah,$^4$ M.~Griffin,$^5$ C.~Lacey,$^3$ E.~Le Floc'h,$^6$ \newauthor A.~Papadopoulos,$^7$ M.~Sargent,$^1$ J.M.~Scudder,$^1$ M.~Vaccari,$^{8,9}$ \newauthor I. ~Valtchanov,$^{10}$ L.~Wang$^{11}$\\
$^1$Astronomy Centre, Department of Physics and Astronomy, University of Sussex, Falmer, Brighton BN1 9QH, UK\\
$^2$Department of Statistics, University of Warwick, Coventry, CV4 7AL, UK\\
$^3$Institute for Computational Cosmology, Department of Physics, University of Durham, South Road, Durham DH1 3LE, UK\\
$^4$Virginia Polytechnic Institute \& State University, Department of Physics, MC 0435, 910 Drillfield Drive, Blacksburg, VA 24061, USA\\
$^5$Cardiff School of Physics and Astronomy, Cardiff University, Queens Buildings, The Parade, Cardiff CF24 3AA\\
$^6$Service d'Astrophysique, CEA-Saclay, Orme des Merisiers, Bat.709, 91191 Gif-sur-Yvette, France\\
$^7$European University Cyprus, 6, Diogenis Str., Engomi, P.O. Box: 22006, 1516 Nicosia-Cyprus\\
$^8$Department of Physics and Astronomy, University of the Western Cape, Robert Sobukwe Road, 7535 Bellville, Cape Town, South Africa\\
$^9$INAF - Istituto di Radioastronomia, via Gobetti 101, 40129 Bologna, Italy\\
$^{10}$Herschel Science Centre, ESAC, ESA, Villafranca del Castillo, Spain\\
$^{11}$SRON Netherlands Institute for Space Research, Landleven 12, 9747 AD, Groningen, The Netherlands}
\date{Released 2002 Xxxxx XX}

\pagerange{\pageref{firstpage}--\pageref{lastpage}} \pubyear{2002}

\def\LaTeX{L\kern-.36em\raise.3ex\hbox{a}\kern-.15em
    T\kern-.1667em\lower.7ex\hbox{E}\kern-.125emX}

\graphicspath{}
\begin{document}

\label{firstpage}
\maketitle

\begin{abstract}

We have developed a new prior-based source extraction tool, \textsc{XID+}, to carry out photometry in the \emph{Herschel} SPIRE maps at the positions of known sources. \textsc{XID+} is developed using a probabilistic Bayesian framework which provides a natural framework in which to include prior information, and uses the Bayesian inference tool \emph{Stan} to obtain the full posterior probability distribution on flux estimates. In this paper, we discuss the details of \textsc{XID+} and demonstrate the basic capabilities and performance by running it on simulated SPIRE maps resembling the COSMOS field, and comparing to the current prior-based source extraction tool \textsc{DESPHOT}. We show that not only does \textsc{XID+} perform better on metrics such as flux accuracy and flux uncertainty accuracy, we illustrate how obtaining the posterior probability distribution can help overcome some of the issues inherent with maximum likelihood based source extraction routines. We run \textsc{XID+} on the COSMOS SPIRE maps from HerMES, using a 24 $\mathrm{\mu m}$ catalogue as a prior and show the marginalised SPIRE colour-colour plot and marginalised contribution to the cosmic infrared background at the SPIRE wavelengths. 
\textsc{XID+} is a core tool arising from the \emph{Herschel} Extragalactic Legacy Project (HELP) and we discuss how additional work within HELP providing prior information on fluxes can and will be utilised. The software is available at \url{https://github.com/H-E-L-P/XID_plus}. We also provide the data product for COSMOS.  We believe this is the first time that the full posterior probability of galaxy photometry has been provided as a data product. 
\end{abstract}

\begin{keywords}
galaxies: statistics -- infrared: galaxies
\end{keywords}
\section{Introduction}
Ever since the discovery of the far-infrared background by the \emph{Cosmic Background Explorer} COBE; \citep{Puget:1996}, surveys have aimed to observe and detect the sources responsible. Most of those sources are galaxies, with the far-infrared emission coming from dust. 

While ground-based observatories such as SCUBA \citep{SCUBA}, and more recently SCUBA-2 \citep{SCUBA2} and ALMA can make use of infrared atmospheric transmission windows to observe at the tail of the cosmic infrared background (CIB), only space-borne facilities can observe at the peak ($\approx140\mathrm{\mu m}$). The first infrared space telescope, the InfraRed Astronomical Satellite (IRAS; \cite{Neugebauer:1984}), observed the whole sky in four bands centred at 12, 25, 60 and 100 $\mu m$ and revealed new populations of galaxies which were optically faint but luminous in the infrared \citep{Soifer:1984}.

While the Infrared Space Observatory, ISO; \cite{Kessler:1996} and the Spitzer Space Telescope \citep{Werner:2004} have provided deep near and mid-infrared photometry over small fields, other smaller space-borne facilities such as AKARI \citep{Murakami:2007} and the Wide-field Infrared Survey Explorer, WISE; \cite{Wright:2010} have surveyed the entire sky at mid to far-infrared and near to mid infrared wavelengths respectively. The most recent advance in infrared astronomy has been made with the ESA \emph{Herschel} Space Observatory \citep{Pilbratt:2010}. Photometry from the Photoconductor Array Camera and Spectrometer \citep[PACS;][]{Poglitsch:2010} and Spectral and Photometric Imaging Receiver \citep[SPIRE;][]{Griffin:2010} have given us an unprecedented view of the far-infrared Universe by providing observations that measure across the peak of the far-infrared background and at greater sensitivity and resolution than has been achieved previously at these wavelengths, thereby definitively setting the origin of the CIB.

With surveys such as the \emph{Herschel} Multi-Tiered Extragalactic Survey, HerMES; \cite{Oliver:2012} and the \emph{Herschel} ATLAS survey, H-ATLAS; \cite{Eales:2010}, over 1000 square degrees of the sky has been observed by the SPIRE instrument. However, due to the relatively large beam size of the SPIRE, and the galaxy density ($\approx 30$ per SPIRE beam for optical sources with B $<$ 28), multiple galaxies can be located within the SPIRE beam. This is referred to as the problem of source confusion.

To obtain accurate photometry from the SPIRE maps, overcoming the source confusion problem is essential. One way to solve the problem is to use prior information to accurately distribute the flux in the SPIRE maps to the underlying astronomical objects. For example, if we know the location of a galaxy to a reasonable tolerance (e.g. from an optical image where resolution is better), we may expect a galaxy to be found in the SPIRE maps at the same location.

Several techniques have been developed that utilise the positions of sources detected at other wavelengths, usually 24 $\mathrm{\mu m}$ and 1.4 GHz, to disentangle the various contributions from discrete sources to the SPIRE flux in a given beam element \citep[e.g.][]{Roseboom:2010, Roseboom:2011, Chapin:2011}. This process is made possible by the strong correlation between the 24-$\mathrm{\mu m}$ and 1.4-GHz populations and those observed at far-IR wavelengths; $>$80 per cent of the cosmic IR background at SPIRE wavelengths can be accounted for by 24-$\mathrm{\mu m}$ sources with S24 $>$ 25 $\mathrm{\mu Jy}$ \citep[e.g.][]{Marsden:2009, Pascale:2009, Elbaz:2010, Bethermin:2012}, while the strong correlation between the far-IR and radio luminosity is known to hold across a wide range in redshift and luminosity \citep[e.g.][]{Ivison:2010}. Up to the present day, most of these techniques have used a maximum likelihood optimisation approach, which suffers from two major issues. The first is that variance and co-variance of source fluxes can not be properly estimated. The second is that of overfitting when many of the input sources are intrinsically faint. The list driven algorithm developed for HerMES \citep[DESPHOT][]{Roseboom:2011,Wang:2014} tried to overcome this by using the non-negative weighted LASSO algorithm \citep{Tibshirani:1996, Zou:2006, terBraak:2010}, a shrinkage and selection method which introduces an additional penalty term in an attempt to reduce the number of sources needed to fit the map. However, when multiple sources are located close-by (i.e. within the SPIRE beam), the method has been found to wrongly assign all the flux to one source. 

The solution to both of these problems is to fully explore the posterior probability distribution with Bayesian inference techniques such as Markov Chain Monte Carlo (MCMC) methods. By fully exploring the posterior, the variance and covariance between sources can be properly estimated. Also, by considering the covariance between sources (i.e. how the flux of sources affect each other), the probability of sources being very faint or bright is taken into account, removing the need for methods such as LASSO. 

Up until the present day, use of MCMC techniques has been computationally unfeasible. However, advances in computational technology and algorithms such as Hamiltonian Monte Carlo now make this sort of approach a viable alternative, as demonstrated by \cite{Safarzadeh:2015}, who used an MCMC based approach to fit PACS simulated maps.
 
As part of the \emph{Herschel Extragalactic Legacy Project} \citep[HELP;][]{Vaccari:2015, Oliver:2016}, we have developed an alternative prior based approach for source extraction in confusion-dominated maps. Our new method, \textsc{XID+}, is built upon a Bayesian probabilistic framework which provides a natural way in which to introduce additional prior information. By using the Bayesian inference tool, \citep[\emph{Stan}, ][]{pystan-software:2015, stan-software:2015} to sample the full posterior distribution, we are also able to provide more accurate flux density error estimates, whilst avoiding some of the issues associated with the maximum likelihood and LASSO fitting approach used by \textsc{DESPHOT}. In this paper, we show that \textsc{XID+} outperforms \textsc{DESPHOT} when using just positional information. In Section \ref{sec:XID+} we discuss the algorithm, and show how the software performs on simulated SPIRE maps in Section \ref{sec:sims}. In Section \ref{sec:COSMOS} we apply \textsc{XID+} on the HerMES COSMOS SPIRE maps, using a 24 $\mathrm{\mu m}$ catalogue as a prior and show the resulting marginalised SPIRE colour-colour plot and contribution to the cosmic infrared background. We discuss how \textsc{XID+} can make use of flux prior information, delivered by the HELP project (Hurley et al. in prep) in Section \ref{sec:disc} and make final conclusions in \ref{sec:conc}.
 
\section{\textsc{XID+} Algorithm}\label{sec:XID+}
The basic goal of \textsc{XID+} is to use the SPIRE maps to infer the likely SPIRE flux of sources we already know about. Bayesian inference is well suited to these requirements. It allows the use of prior information and provides a posterior distribution of the parameter(s) after taking into account the observed data.  

We also want to provide a framework to do science directly with the maps rather than adding the additional step of first creating catalogues, which in essence is a form of lossy data compression.

We therefore adopt a Bayesian probabilistic modelling approach for our XID+ algorithm. It aims to:
\begin{itemize}
\item map out the posterior rather than the traditional maximum likelihood point estimate, thereby providing a full account of the flux uncertainty; 
\item extend the use of prior information beyond just using positional information about sources.
\end{itemize}

In the following section, we describe our \textsc{XID+} algorithm. As this algorithm builds upon knowledge gained from the original XID (a.k.a \textsc{DESPHOT}) algorithm used by HerMES \citep{Roseboom:2010, Roseboom:2011, Wang:2014}, we describe XID+ in the context of how it differs from \textsc{DESPHOT}. 


\subsection{Basic Model}
Our data ($\mathbf{d}$) are maps with $n_1 \times n_2 = M$ pixels. Our model assumes the maps are formed from $S$ known sources, with flux density $\mathbf{f}$ and a background term accounting for unknown sources. The point response function (PRF) tells us the contribution each source makes to each pixel in the map and is assumed to be a Gaussian, with full-width half-maximum (FWHM) of 18.15, 25.15 and 36.3 arcsec for 250, 350 and 500 $\mathrm{\mu m}$ respectively \citep{Griffin:2010}. Our map can therefore be described as follows:

\begin{equation}
\mathbf{d} = \sum\limits_{i=1}^S \mathbf{P f_i} + N(0,\Sigma_{instrumental}) + N(B,\Sigma_{confusion})
\label{eq:map}
\end{equation}
where $\mathbf{d}$ is our model of the map, $\mathbf{P}$ is the PRF, $f_i$ is the flux density for source $i$ and two independent noise terms, one for instrumental noise, the other for confusion noise which we model as Gaussian fluctuations about $B$, a global background.

We can rewrite the above equation in the linear form:
\begin{equation}
\mathbf{d} = \mathbf{Af}
\label{eq:map2}
\end{equation}
Where $\mathbf{d}$ is flattened ta a vector with $M$ pixels, $A$ is a sparse $M \times S$ matrix pointing matrix. For SPIRE, the pointing matrix is calculated by taking the Gaussian PRF for each band at a 1 arcsecond pixel scale, then centering it on the position for each source and carrying out a nearest neighbour interpolation to establish the contribution each source makes to each pixel in the map.

As instrumental and confusion noise are independent, we can combine the two noise terms into one covariance matrix such that $\mathbf{N_d} = \Sigma_{instrumental}+\Sigma_{confusion}$. Confusion noise will be correlated across nearby pixels due to the PRF and across the three SPIRE bands. Taking these correlations into account requires the full $M \times M$ covariance matrix, which vastly increases computational time. For simplicity we currently ignore the correlations and assume the confusion noise is constant across the map. The covariance matrix becomes a diagonal matrix i.e. $\mathbf{N_{d,ii}} =\sigma_{inst.,ii}^2+\sigma_{conf.}^2$. 

 We can now define the likelihood as the Gaussian probability function for the data given the flux densities
\begin{equation}
L = p(\mathbf{d}|\mathbf{f}) \propto |\mathbf{N_d}|^{-1/2} \exp\big\{ -\frac{1}{2}(\mathbf{d}-\mathbf{Af})^T\mathbf{N_d}^{-1}(\mathbf{d}-\mathbf{Af})\big\}\label{eq:likelihood}
\end{equation}
The maximum likelihood solution to this equation can be found by setting $\chi = (\mathbf{d}-\mathbf{Af})^T\mathbf{N_d}^{-1}(\mathbf{d}-\mathbf{Af})$, finding the minimum and rearranging such that:

\begin{equation}
\mathbf{f}=(\mathbf{A^TN_d^{-1}A})^{-1}\mathbf{A^TN_d^{-1}d}\label{eq:mlm}
\end{equation}

Equation \ref{eq:mlm} can be solved directly, either by brute-force matrix inversion or via other linear methods. As discussed in \cite{Roseboom:2010, Roseboom:2011, Wang:2014}, linear approaches ignore prior knowledge that fluxes cannot have negative flux density.  They are also incapable of discriminating between real and spurious soltuions, which can result in overfitting. To overcome these issues, \cite{Roseboom:2011} used the non-negative weighted LASSO algorithm \citep{Tibshirani:1996, Zou:2006, terBraak:2010}.

LASSO is a shrinkage and selection method for linear regression and works by treating sources either as `inactive' with flux density set to zero, or `active'. It switches sources on one at a time, with the order determined by reduction in chi-squared gained by turning them on. The process continues until some tolerance is reached.

%

For \textsc{XID+}, we want to map out the entire posterior, $p(\mathbf{f}|\mathbf{d})$, rather than find the maximum likelihood solution. This has the benefit that it gives us more complete information on how certain we are about the predicted fluxes. The posterior can be defined as:
\begin{equation}
p(\mathbf{f}|\mathbf{d}) \propto p(\mathbf{d}|\mathbf{f}) \times p(\mathbf{f})
\end{equation}
where $p(\mathbf{d}|\mathbf{f})$ is our likelihood, defined in equation \ref{eq:likelihood} and $p(\mathbf{f})$ is our prior on the fluxes. For our simplest model, we use a uniform distribution for $p(\mathbf{f})$, with an upper bound of 1000 mJy and lower bound of 0.01mJy.

\begin{figure}
\includegraphics[width=8.5cm]{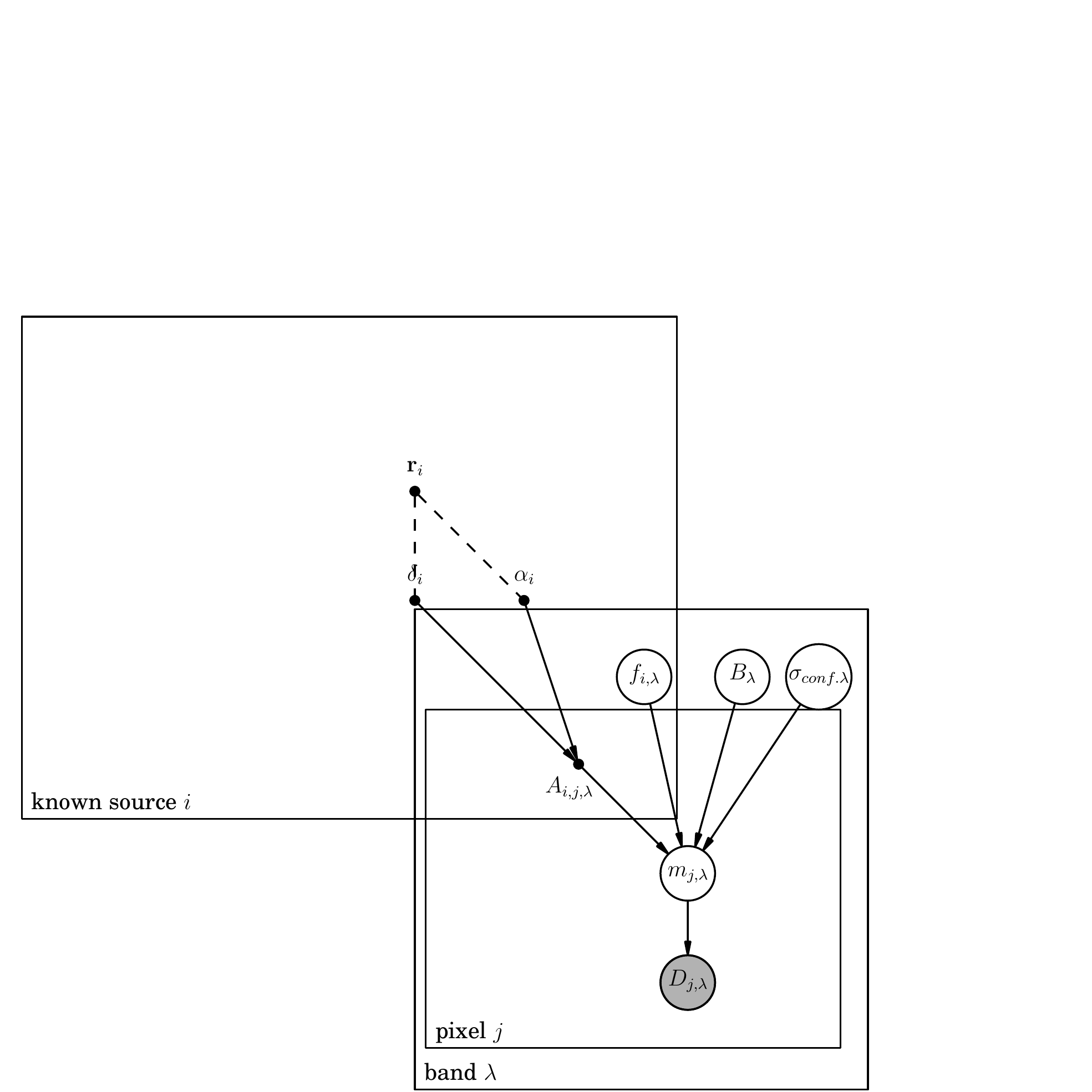}
\caption{Our probabilistic model for\textsc{XID+}. Boxes represent repeated dimensions, open circles as variables, dots as deterministic (or fixed) variables. Created with DAFT (\url{http://daft-pgm.org/})}\label{fig:graph_mod_xid+}
\end{figure}

In our probabilistic framework, we can illustrate our model for the map, defined in equation \ref{eq:map2} via a probabilistic graphical model (PGM). Figure \ref{fig:graph_mod_xid+} shows a plate diagram \citep{Bishop:2006} for our PGM of the basic XID+ model, where boxes  indicate repeated values such as source ($i$), pixel ($j$) and band ($\lambda$). Open circles correspond to random variables and dots are deterministic (or fixed) variables, with their relative positions in the boxes indicating what indices they repeat over. For our simplest model, the positional vector of sources ($\mathbf{r_i}$) can be described by sky co-ordinates, $\alpha_i$ and $\delta_i$ and are treated as deterministic (i.e. known). \footnote{In reality, positional information is often uncertain. However, these uncertainties are relatively small in comparison to the SPIRE beams, for example, the 24 micron source catalogues in \citep{LeFLoch:2009} are accurate to $\approx 2''$ in respect to $K$ band catalogues, this corresponds to a ninth of the FWHM of the SPIRE beam at 250 microns}. The PRF is assumed to be a Gaussian, with full-width half-maximum (FWHM) of 18.15, 25.15 and 36.3 arcsec for 250, 350 and 500 $\mathrm{\mu m}$ respectively \citep{Griffin:2010}. Both these deterministic variables are used to make the pointing matrix $A_{i,j,\lambda}$ which gives the contribution source $i$ makes to each pixel $j$ in the map at wavelength $\lambda$. Each source has its own flux $f_{i,\lambda}$ which is a random variable. By multiplying $f$, $A$ for all sources and pixels, and adding our global estimate for the background $B$, we can make our model for the map, $m$, which we can compare to the data $D$. 
 
\subsubsection{Stan}
Now that we have our probabilistic model, we need to sample from it to obtain the posterior. We use the Bayesian inference tool, \textit{Stan}, which is `a probabilistic programming language implementing full Bayesian statistical inference with MCMC sampling'. \textit{Stan} uses the adaptive Hamiltonian Monte Carlo (HMC) No-U-Turn Sampler (NUTS) of \cite{Hoffman:2013} to efficiently sample from the posterior. It does this by using the gradient information, allowing fast traversing of high dimensional and highly correlated joint posterior distributions. 

\textit{Stan} has its own modelling language, in which one constructs probabilistic models. Our model for \textit{Stan} can be found in Appendix A.
\subsubsection{Estimating Convergence}\label{sec:conv}
As with all MCMC routines, one needs to run enough chains and run them long enough to be confident the global minimum has been found and that it has been thoroughly sampled. 

As default, we run four separate chains from different initial positions in parameter space. We also discard the first half of the chain as `warm up' to ensure the chains have converged to the posterior distribution. We then assess the convergence of each parameter by comparing the variation between and within chains using the diagnostics described in \cite{BDA3} which can be summarised as follows: Each chain is split in two and the between-chain ($BC$) and within-chain ($WC$) variance is calculated. $BC$ and $WC$ are then used to calculate the marginal posterior variance. This in turn can be used to estimate the potential scale reduction $\hat{R}$, which reduces to 1 as the number of iterations tends to infinity. An $\hat{R}$ value $> 1.2$ suggests chains require more samples. We provide $\hat{R}$ for each parameter.

Due to the nature of MCMC, samples from MCMC routines are correlated. Inference from correlated samples is less precise than from the same number of independent draws. In order to check there are enough independent draws we estimate the effective number of samples $\hat{n_{eff}}$, defined in \cite{BDA3}. We require $\hat{n_{eff}}$ to be 10 times the number of chains and provide the estimate for each parameter.

\subsection{Map segmentation}
The survey fields in HELP vary in size from 0.3 to 290 square degrees. Ideally, source photometry and background estimation would be done on the full image. In practice this will be computationally unfeasible. \textsc{DESPHOT} segmented the map by locating islands of high SNR pixels enclosed by low SNR pixels.

%
%
We adopt a simpler tiling scheme that splits map data into equal area diamonds based on the Hierarchical Equal Area isoLatitude Pixelization of a sphere (HEALPix). The resolution of the pixels are determined by the HEALPix level, with default for \textsc{XID+} set at 11 which corresponds to $\approx 1.718'$. When fitting each tile, the perimeter being fitted is extended by one HEALPix pixel with a resolution which is 2 levels higher (i.e. default is level 13 with a resolution of $\approx 25.77'$) such that all sources that could foreseeably contribute to sources within the HEALPix pixel of interest are taken into account. To give an example of the dimensionality, our fit to a mock simulation (described in the following section),  uses a HEALPix tile at order 9, fitting over 600 sources on average to  $\approx$10,500 pixels at 250 microns, $\approx$5500 pixels at 350 microns and $\approx$2500 pixels at 500 microns. That means for each tile we are fitting over 1800 parameters simultaneously.


The choice of HEALPix pixel size affects the computational time of \textsc{XID+}. The required CPU time is found to scale linearly with the amount of data (i.e. number of image pixels).

\subsection{Uncertainties and Covariances}
\textsc{DESPHOT} provides an estimate of the covariance matrix associated with the fluxes ($\mathbf{N_f}$) from $(\mathbf{A^TN_d^{-1}A})^{-1}$. Due to the Cramer-Rao inequality, this estimate is a lower limit. It also assumes the \textsc{DESPHOT} algorithm is linear, which is not strictly true having introduced LASSO and non-negative priors. As a result, the uncertainties are unreliable. For \textsc{XID+}, we have the full posterior, allowing the true variance to be properly characterised. This not only gives us a better estimate for marginalised uncertainty for each source, but it also provides the covariance information between sources (as seen in Figure \ref{fig:corr}). 
%


\begin{figure*}
\centering 
\subfloat[$250\mathrm{\mu m}$]{\includegraphics[width=5.5cm,page={4}]{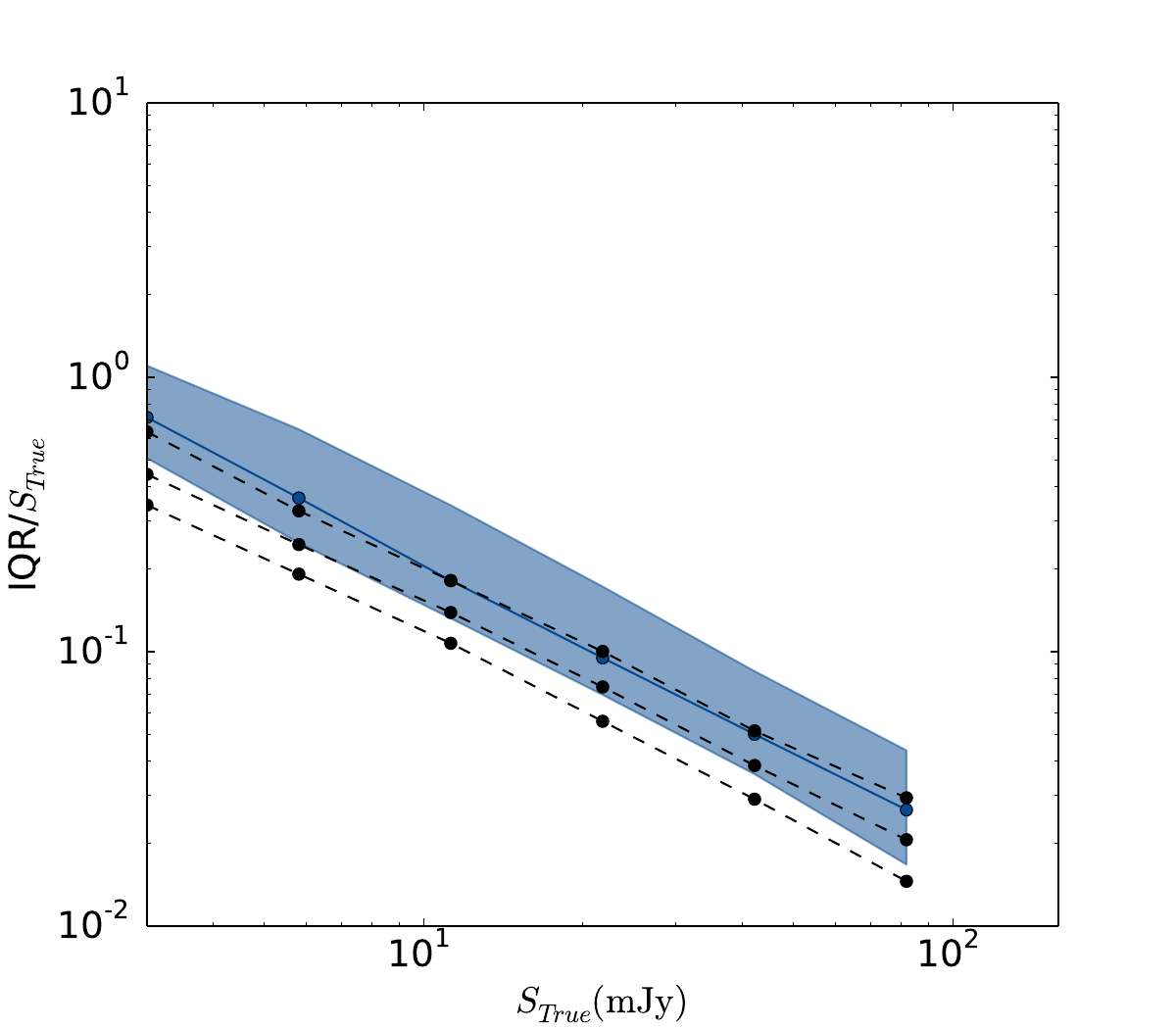}}
\subfloat[$350\mathrm{\mu m}$]{\includegraphics[width=5.5cm,page={5}]{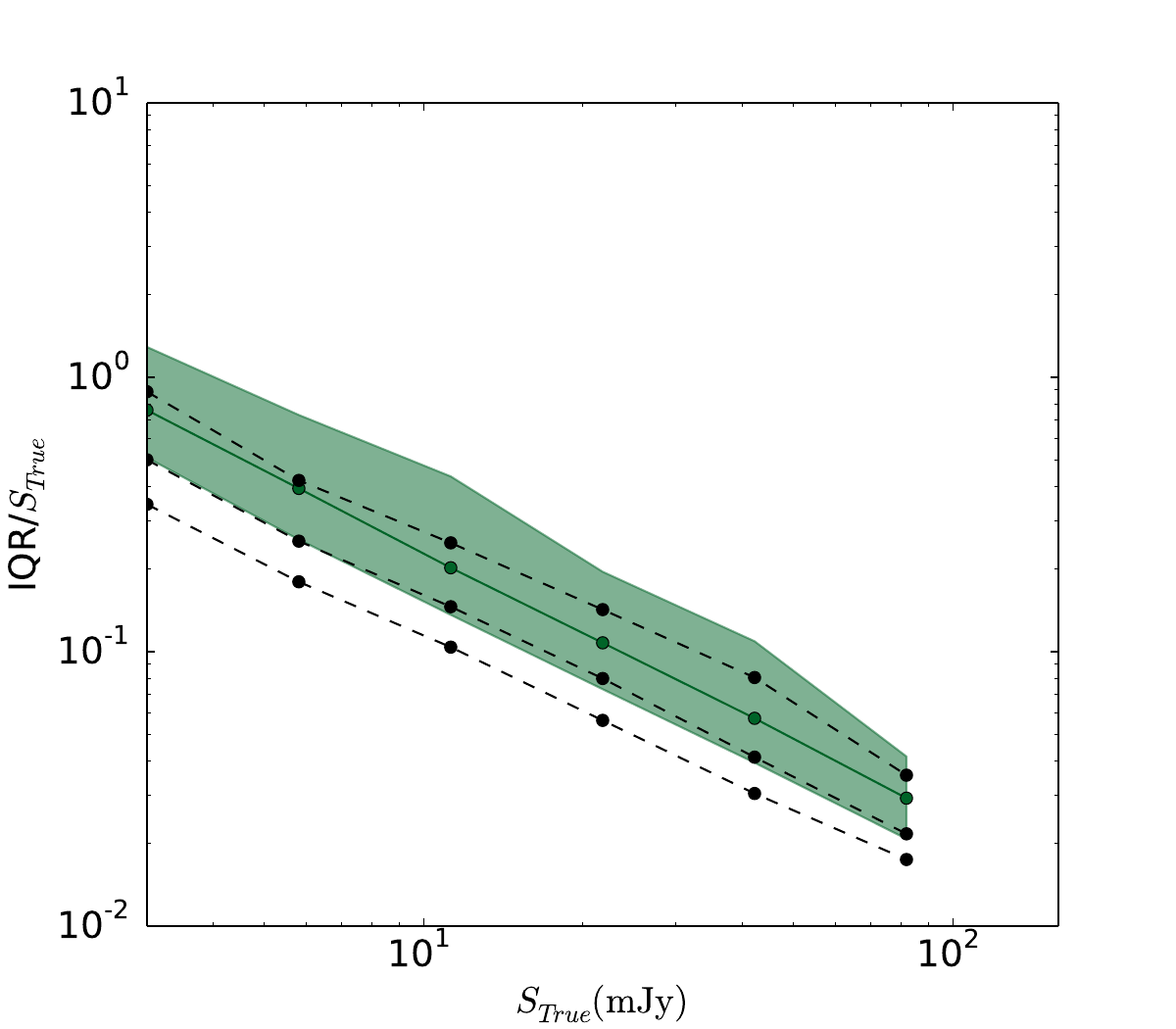}}
\subfloat[$500\mathrm{\mu m}$]{\includegraphics[width=5.5cm,page={6}]{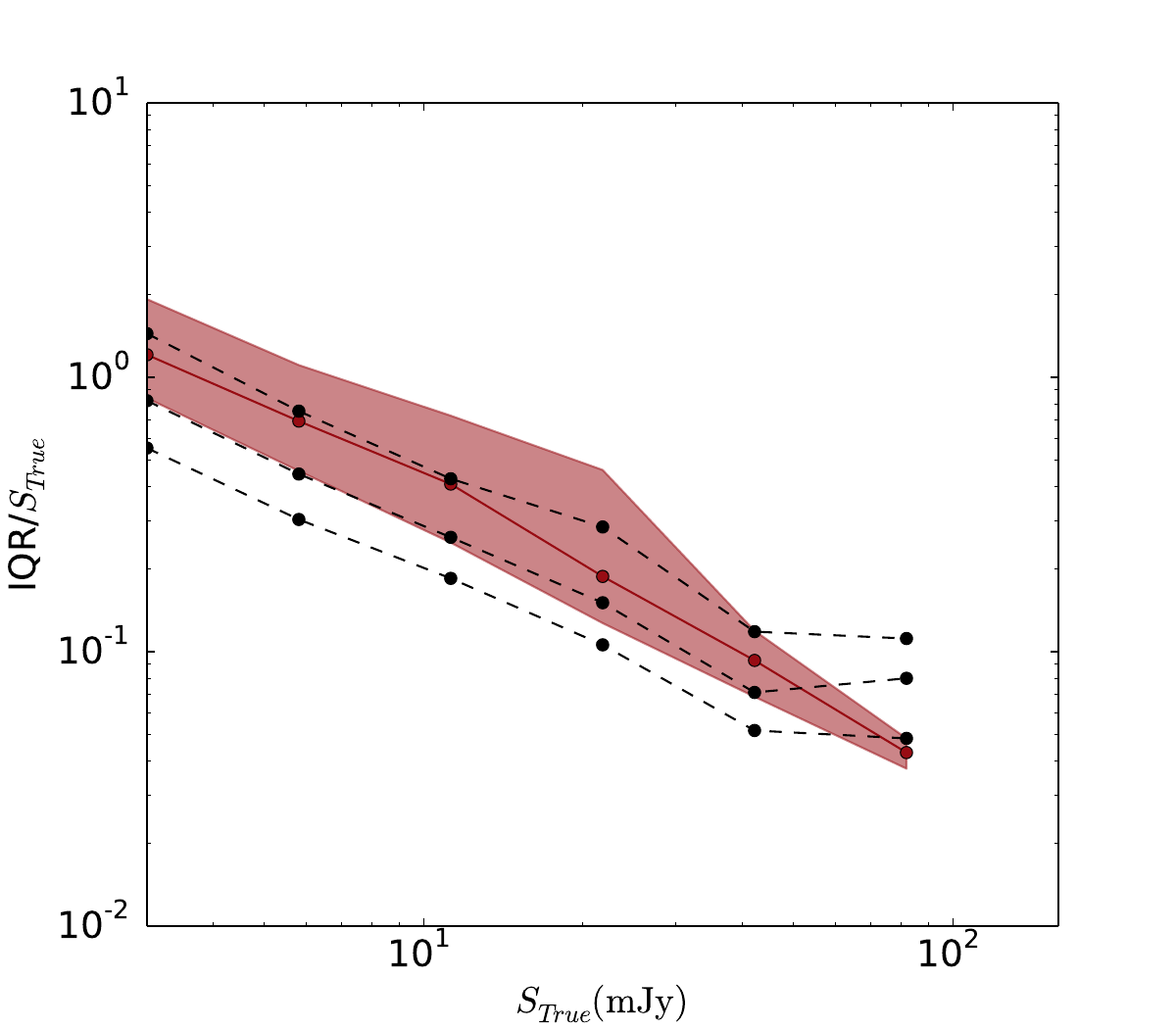}}
\caption{The 16th, 50th and 84th percentiles for inverse precision, or IQR of \textsc{XID+} (coloured line and shaded region) and \textsc{DESPHOT}  (black dashed lines), as a function of true flux, for the 250 (blue), 350 (green) and 500 (red) $\mathrm{\mu m}$ SPIRE bands. }\label{fig:precision}
\end{figure*}

\begin{figure*}
\centering 
\subfloat[$250\mathrm{\mu m}$]{\includegraphics[width=5.5cm,page={7}]{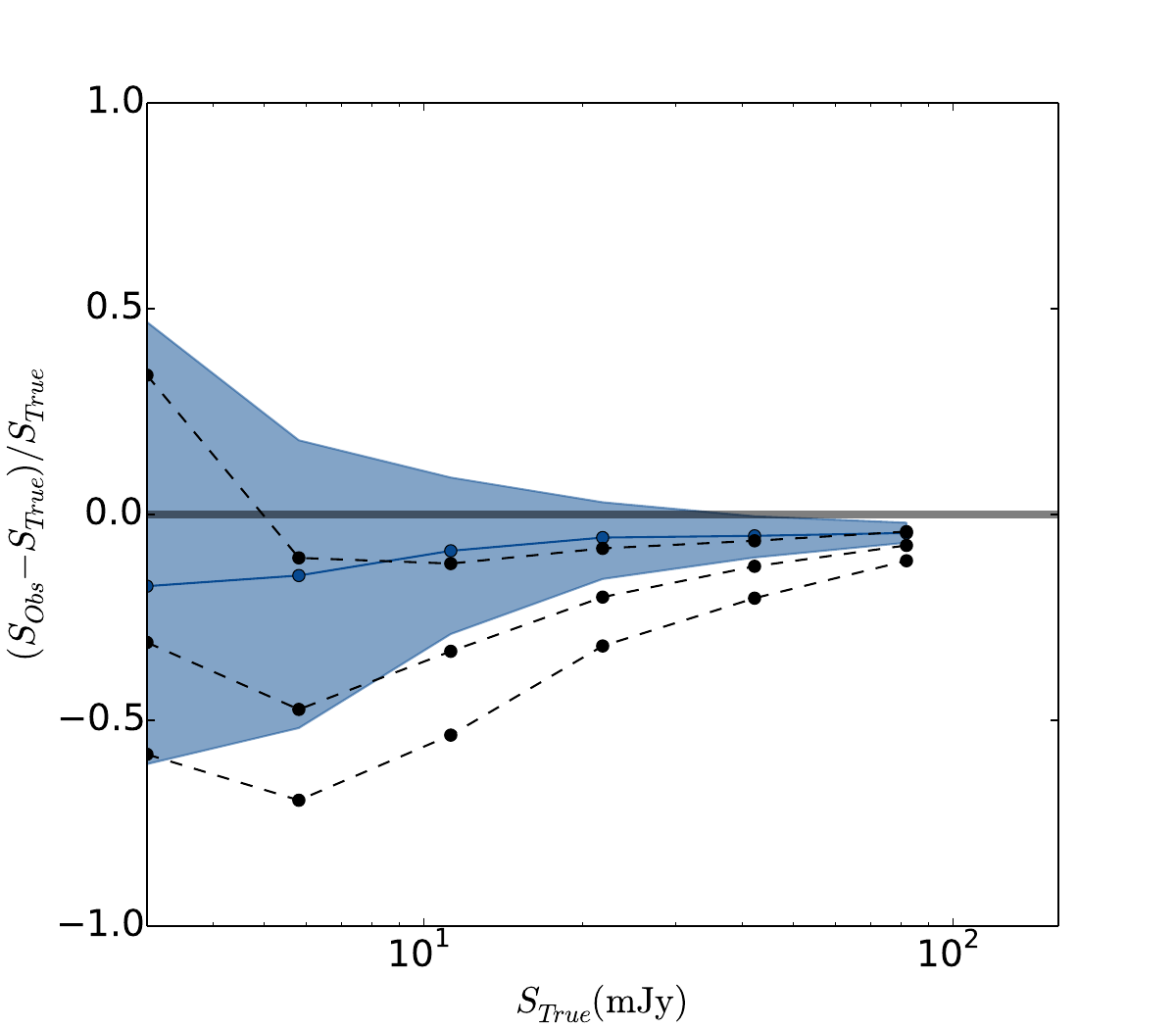}}
\subfloat[$350\mathrm{\mu m}$]{\includegraphics[width=5.5cm,page={8}]{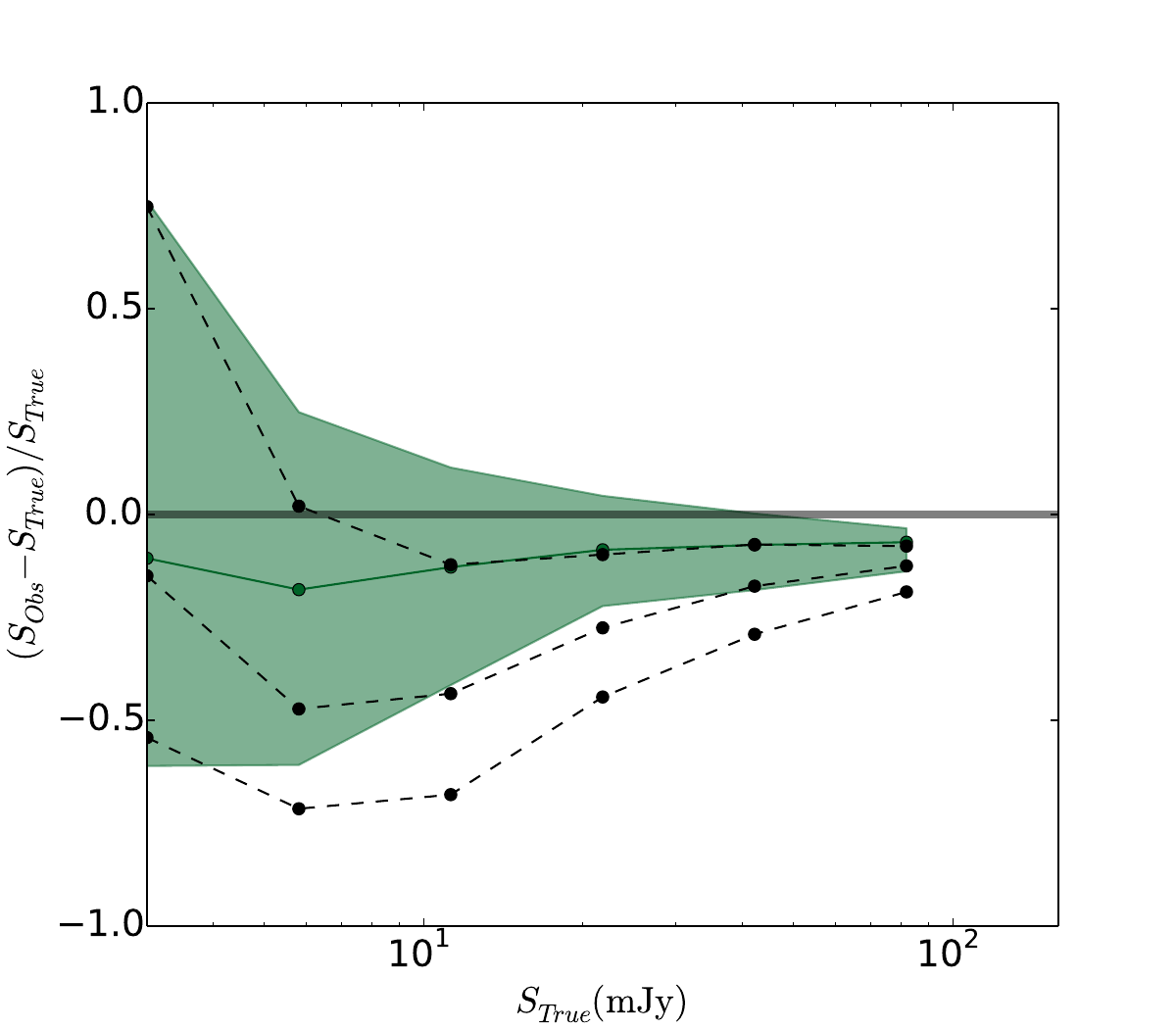}}
\subfloat[$500\mathrm{\mu m}$]{\includegraphics[width=5.5cm,page={9}]{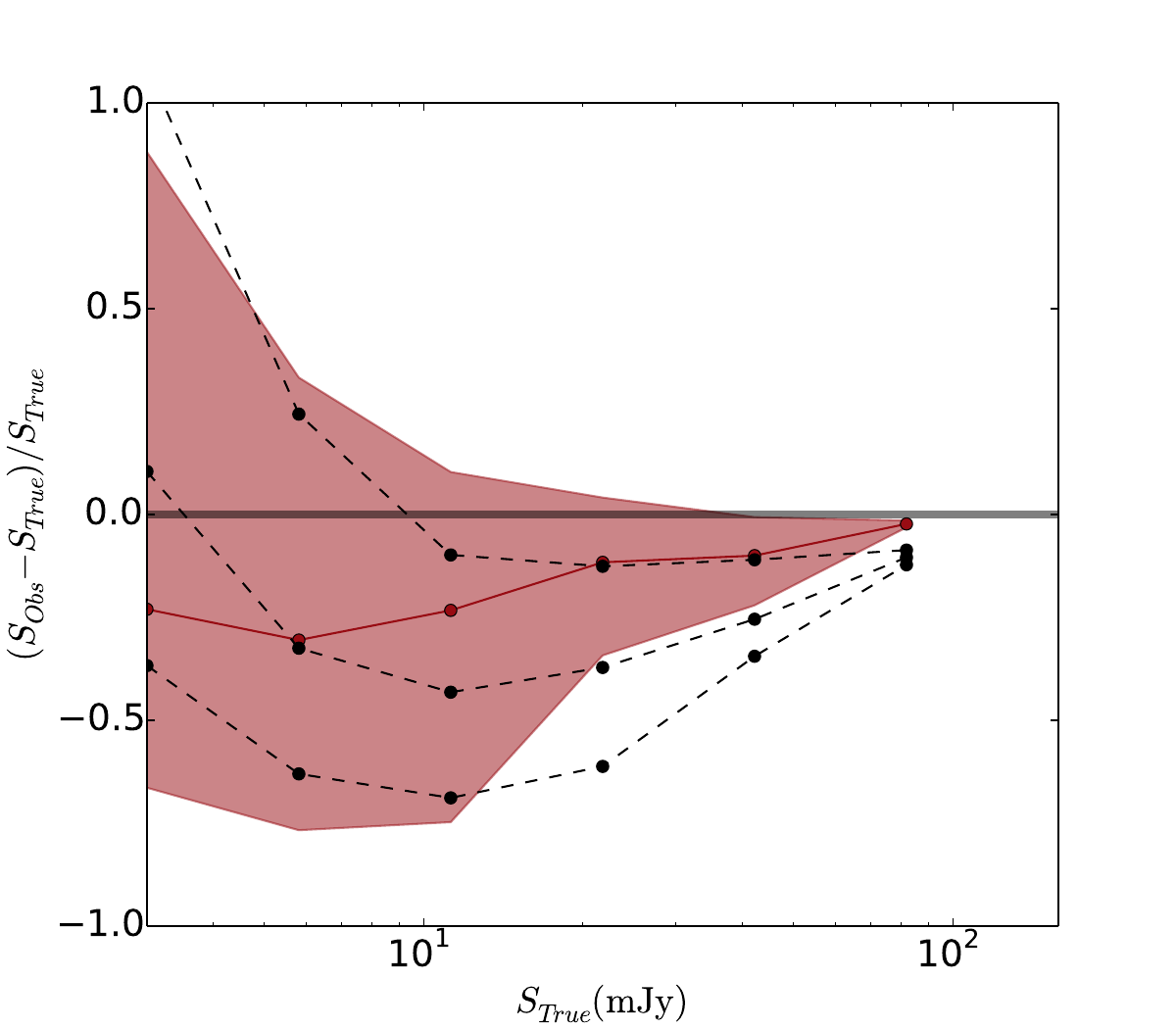}}
\caption{The 16th, 50th and 84th percentiles for flux accuracy of \textsc{XID+} (coloured line and shaded region) and \textsc{DESPHOT} (black dashed lines), as a function of true flux, for the 250 (blue), 350 (green) and 500 (red) $\mathrm{\mu m}$ SPIRE bands. A horizontal thick line is shown at zero for clarity.}\label{fig:accuracy}
\end{figure*}

\begin{figure*}
\centering 
\subfloat[$250\mathrm{\mu m}$]{\includegraphics[width=5.5cm,page={1}]{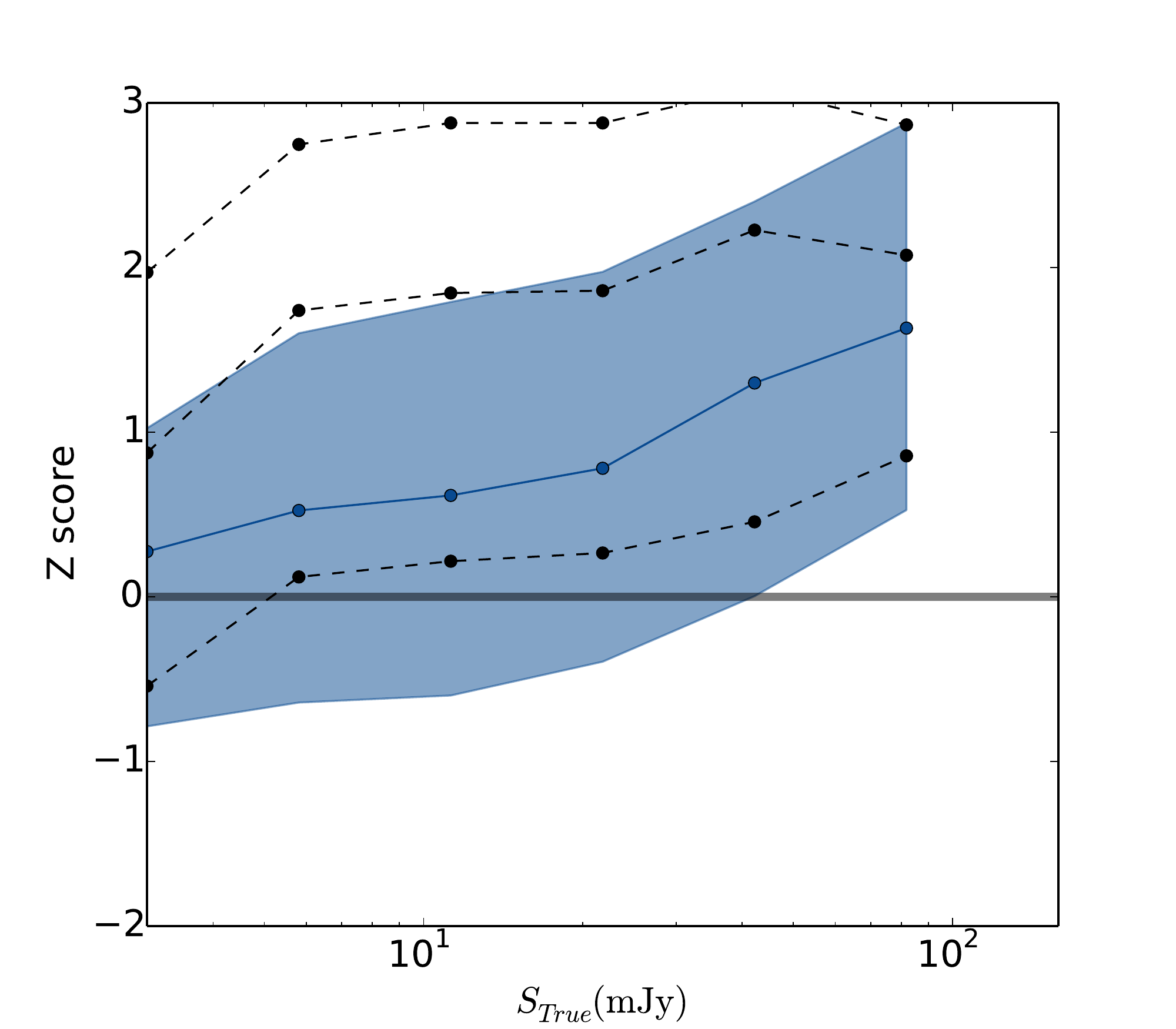}}
\subfloat[$350\mathrm{\mu m}$]{\includegraphics[width=5.5cm,page={2}]{metrics_both.pdf}}
\subfloat[$500\mathrm{\mu m}$]{\includegraphics[width=5.5cm,page={3}]{metrics_both.pdf}}
\caption{The 16th, 50th and 84th percentiles for Z score, or flux density error of \textsc{XID+} (coloured line and shaded region) and \textsc{DESPHOT}  (black dashed lines), as a function of true flux, for the 250 (blue), 350 (green) and 500 (red) $\mathrm{\mu m}$ SPIRE bands. A horizontal thick line is shown at zero for clarity.}\label{fig:zscore}
\end{figure*}

\section{Simulations}\label{sec:sims}
In order to test and quantify the performance of XID+, we use simulated SPIRE maps of the COSMOS field, a good example of a deep map i.e. where confusion noise ($\sigma_{conf.}$) is much larger than instrumental noise ($\sigma_{inst.}$). In order to get realistic clustering, we use the mock catalogues from the latest version of the Durham semi-analytic model, \emph{GALFORM} \citep{Lacey:2015,Cowley:2014}. The model is designed to populate Millennium-class, dark matter only, N-body simulations with a WMAP7 cosmology and minimum halo mass of $1.9 \times 10^{10} h^{-1} M_{\odot}$. The dust model is motivated by the radiative transfer code GRASIL \citep{Silva:1998} and can accurately reproduce the predictions for rest-frame wavelengths $\lambda_{rest} > 70 \mathrm{\mu m}$. We pass this mock catalogue through the HerMES mapmaker pipeline \citep[e.g.][]{Levenson:2010,Viero:2013} to generate a mock HerMES observation, with similar noise properties to the observed COSMOS field.

A mock 100 $\mathrm{\mu m}$ input catalogue, similar to what would be expected of a PACS catalogue, is generated by taking the mock catalogue and making a cut at a flux limit of 50 $\mathrm{\mu Jy}$ (similar to that used for a 24 $\mathrm{\mu m}$ input catalogue), giving a total of 64,719 sources over 3.4 square degrees. We use this as our prior input catalogue for both \textsc{XID+} and \textsc{DESPHOT}. In order to compare performance, we look at three measures: precision, flux accuracy, and flux uncertainty accuracy. For \textsc{XID+}, we only consider sources whose output median flux is above 1 $\mathrm{mJy}$. Likewise, with \textsc{DESPHOT}, we only consider sources which have a maximum likelihood flux of greater than 1 $\mathrm{mJy}$. 

\subsection{Flux Precision}
\label{sec:prec}
Precision is a measure of how well the flux is believed to be constrained. For our posterior sample, this relates to the spread of the sample and so we use the interquartile range (75th - 25th percentile; IQR) as our measure of precision. Figure \ref{fig:precision} shows the 16th, 50th and 84th percentile (i.e. median and median $\pm \sigma$) of the IQR for 6 bins in true flux. IQR is normalised as a function of input flux for both \textsc{XID+} and \textsc{DESPHOT}. As one would expect, $\mathrm{IQR}/S_{True}$ decreases as a function of input flux, indicating a higher precision is achieved for the brighter sources. While $250$ and $350\mathrm{\mu m}$ outputs achieve a similar level of precision, the outputs for $500 \mathrm{\mu m}$ do not reach the same level of precision. In comparison to \textsc{DESPHOT}, \textsc{XID+} is marginally less precise for all three bands, though as we show later, \textsc{DESPHOT}s smaller precision comes at a price of severely underestimating uncertianty. 

\subsection{Flux Accuracy}
Flux accuracy is a measure of how far away the estimated flux is from the truth. We use the difference between the median flux estimate from our posterior and the true flux from the simulation, normalised by true flux, as our estimate of flux accuracy. Figure \ref{fig:accuracy} shows how flux accuracy changes as a function of input flux for all three bands. As before, we show the 16th, 50th, and 84th percentile for 6 bins in true flux. For 250 $\mathrm{\mu m}$, $\textsc{XID+}$  reaches an offset smaller than 10\% by $\approx 5 \mathrm{mJy}$, whereas \textsc{DESPHOT} underestimates the flux for all but the very brightest sources. For 350 $\mathrm{\mu m}$ and 500 $\mathrm{\mu m}$, the offset from the truth is less than 10\% by $\approx10 \mathrm{mJy}$, where as \textsc{DESPHOT} continues to underestimate for all but the very brightest sources. There remains a slight offset of $\approx5\%$ for all fluxes. This is likely due to our inability to model the correlated component of the confusion noise.

\subsection{Flux Uncertainty accuracy}\label{sec:unc}
Estimated flux values should be within one sigma of the true value 68.27\% of the time and within 2 sigma 95.45\% of the time. We can quantify how many sigma away the true value is from the median in terms of a Z score. A Z score of 1 corresponding to being 1 sigma above the median. Figure \ref{fig:zscore} shows the flux uncertainty accuracy (or Z score) as a function of input flux. For \textsc{DESPHOT}, uncertainties are assumed to have a normal distribution, truncated at zero. With \textsc{XID+}, we have the full posterior and do not have to make an assumption on the shape of the uncertainty distribution. If we assume the posterior has good frequentist coverage, then we can calculate uncertainty accuracy by taking the percentile at which the true flux value falls within the posterior, and convert the percentile to a corresponding sigma level.

For the 250 $\mathrm{\mu m}$ band, and sources $\lessapprox25\mathrm{m Jy}$, \textsc{XID+} produces a Z score distribution that is slightly above that expected if uncertainties are correctly estimated (i.e. distribution is centred around zero, with width $\approx 1$). Above $25\mathrm{mJy}$, the median Z score increases, indicating that flux uncertainties are being under estimated. In comparison, for all fluxes, the uncertainty distribution from \textsc{DESPHOT} are above 1 and increases to over 2 for the brighter sources. There are also a large number of sources with a Z score greater than 3 (as seen by the higher density in bins at a Z score of 3). This indicates that \textsc{DESPHOT} is a poor estimator with the majority of sources in \textsc{DESPHOT} lying more than 1 $\sigma$ away from their true flux. The flux uncertainty accuracies for 350 $\mathrm{\mu m}$ and 500 $\mathrm{\mu m}$ show a similar behaviour, though not as severe.

\subsection{Convergence}
\begin{figure}
\subfloat{\includegraphics[width=6cm,page={1}]{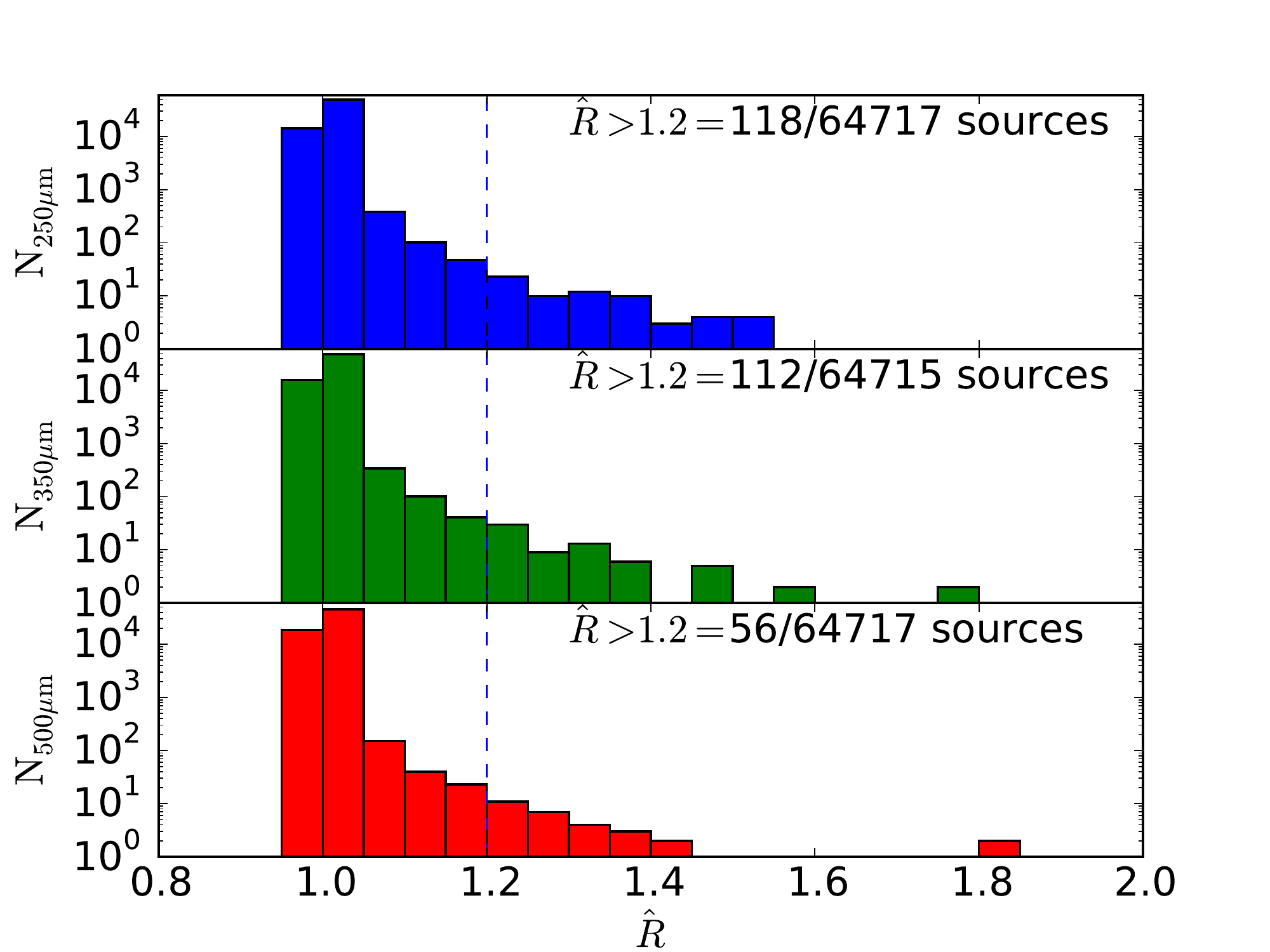}}\\
\subfloat{\includegraphics[width=6cm,page={2}]{convergence_test.pdf}}
\caption{$\hat{R}$ and $\hat{n_{eff}}$ values for all sources fitted in the simulation. The majority of sources have converged and have enough effective samples.}\label{fig:converg}
\end{figure}

As described in Section \ref{sec:conv}, we provide $\hat{R}$ as an estimate of convergence and $\hat{n_{eff}}$ as a measure of independence within the sample. Figures \ref{fig:converg} show the histogram for $\hat{R}$ and $\hat{n_{eff}}$ for the three bands, with our thresholds for the statistics shown by dotted lines. In our fit to the simulated SPIRE maps, we use four chains, each with 1500 iterations (half of which are discarded as warm up). This leads to over 99.99\% of the sources having an $\hat{R}$ and $\hat{n_{eff}}$ within the threshold for all three bands, indicating our solution is well converged. In cases where convergence has not been reached, the number of iterations can be increased.

\subsection{Correlated Sources}
\begin{figure} 
\includegraphics[width=8cm]{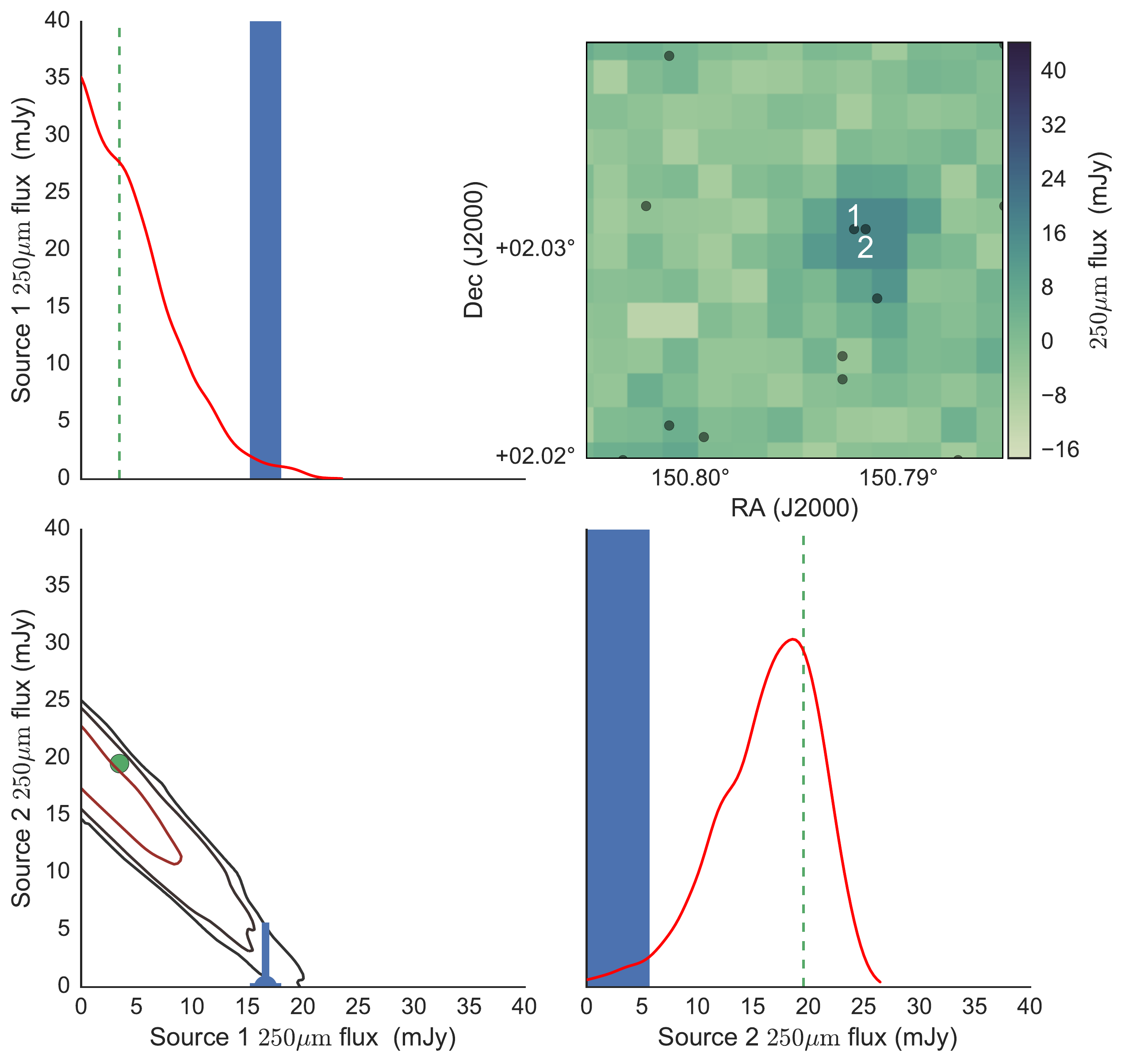}
\caption{Joint and marginalised posterior plot of two correlated sources that are 2'' apart, with one,two and three sigma contours over-plotted. The true flux is shown by the green circle and green dashed lines. \textsc{DESPHOT} (blue error bar and vertical filled region spanning 1 sigma) assigns all the flux to one source which is actually the fainter of the two, where as with \textsc{XID+} we get the full uncertainty information from the posterior.}\label{fig:corr}
\end{figure}

For sources that are close together (i.e. within FWHM of the PRF), the uncertainty on the flux estimates can be correlated. One of the advantages of obtaining the full posterior is that we get a proper estimate of uncertainty and its correlation. This is particularly apparent when comparing flux estimates with \textsc{DESPHOT}, which, by using the LASSO algorithm forces one source to have all the flux and the other nearby source to zero. Figure \ref{fig:corr} shows an example of two sources which are 2'' apart. The 250 $\mathrm{\mu m}$ flux estimate from both \textsc{XID+} and \textsc{DESPHOT} are shown in green and blue respectively. The posterior provided by XID+, fully captures the correlated uncertainty, where as the `winner takes all' approach from \textsc{DESPHOT} clearly fails to estimate the true flux for both sources.
 
\subsection{Defining Detections}\label{sec:dec}
\begin{figure*}
\centering 
\subfloat[$250\mathrm{\mu m}$]{\includegraphics[width=5.5cm,page={1}]{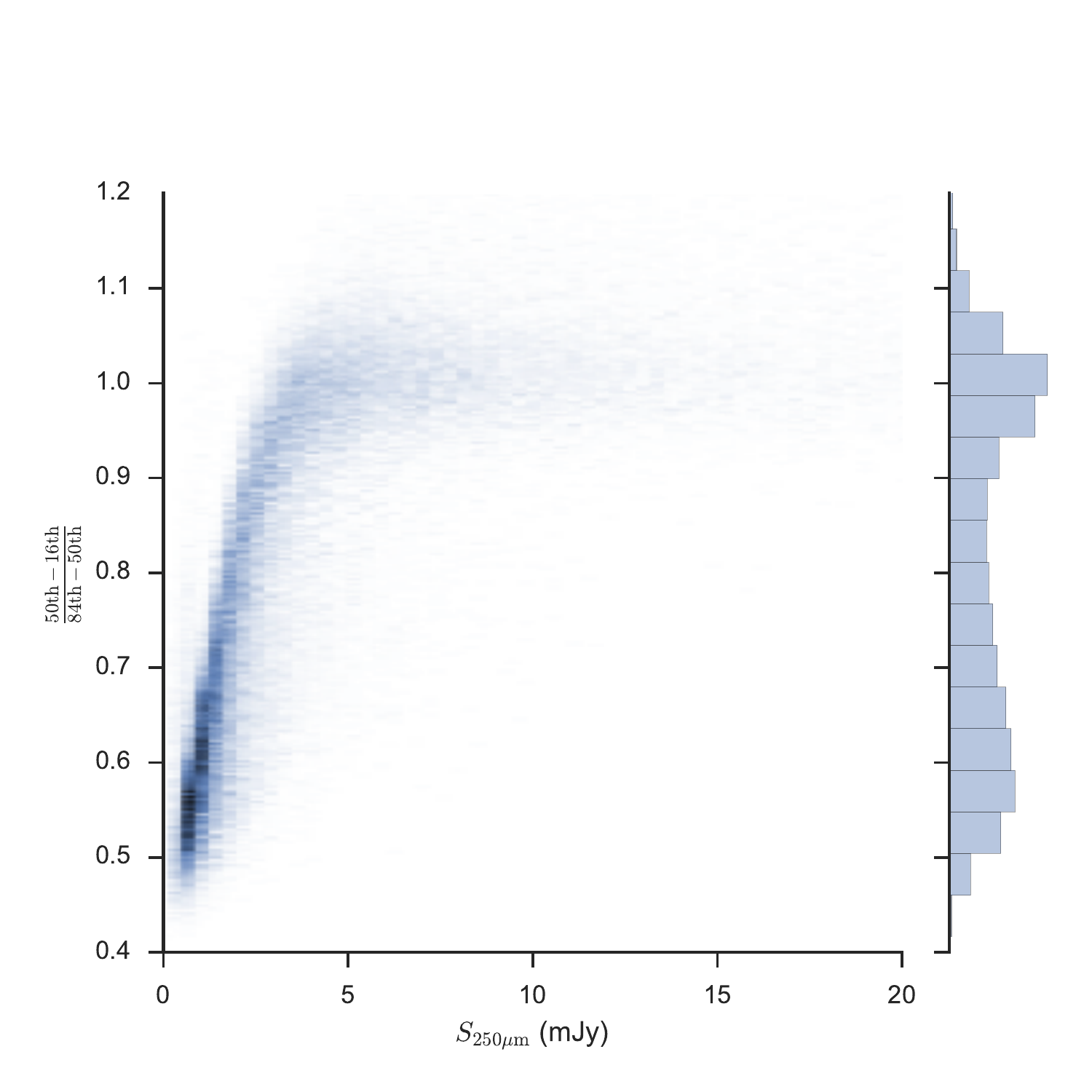}}
\subfloat[$350\mathrm{\mu m}$]{\includegraphics[width=5.5cm,page={2}]{Detections.pdf}}
\subfloat[$500\mathrm{\mu m}$]{\includegraphics[width=5.5cm,page={3}]{Detections.pdf}}
\caption{The $\frac{\mathrm{84th}-\mathrm{50th}}{\mathrm{50th-16th}}$ percentiles of the flux posterior distribution as a function of the $\mathrm{50th}$ percentile for the three SPIRE bands. The posterior is approximately Gaussian when the ratio is around 1. For 250 and 350 $\mathrm{\mu m}$ this occurs at around 4mJy, while for 500 $\mathrm{\mu m}$, it occurs around 6 mJy.}\label{fig:detections}
\end{figure*}

For sources close to or below the noise level of the map, the data will be unable to constrain our model. The flux posterior for these sources will be a subtle combination of the flux prior and an upper limit imposed by the noise level of the map. This results in a non-Gaussian posterior flux distribution for those sources that cannot be constrained. To illustrate how the shape of the posterior flux distribution changes, Figure \ref{fig:detections} shows the $\frac{\mathrm{84th}-\mathrm{50th}}{\mathrm{50th-16th}}$ percentiles of the flux posterior distribution as a function of the $\mathrm{50th}$ percentile for the three SPIRE bands. For 250 and 350 $\mathrm{\mu m}$, uncertainties become Gaussian around 4 mJy, while for 500 $\mathrm{\mu m}$, it occurs around 6 mJy. In principle, the full posterior distribution could be used for all objects, including those not fully constrained by the data. However, Figure \ref{fig:detections} provides a convenient way in which to define the level at which there are robust detections. We note that these limits will change depending on prior list and the maps being fitted.

\subsection{Comparison to Stacking}
To further demonstrate the performance of \textsc{XID+}, we compare the average \textsc{XID+} fluxes for objects grouped into 6 stellar mass and 6 redshift bins, against those from stacking. Stacking takes the position of known sources and for each source, cuts out a thumbnail from a map centred on the source position. By averaging the thumbnail images, one reduces the noise and produces an image of the average galaxy and an average flux estimate for the list of known sources used to make the stack. Figure \ref{fig:stack} in Appendix B shows the 250, 350 and 500 $\mathrm{\mu m}$ stacked, average true flux from the simulation, and average \textsc{XID+} fluxes for 36 different groups of objects. For objects above the detection limit defined in section \ref{sec:dec}, the average flux of the groups from \textsc{XID+} (in red) is able to recover the true average flux (in green) just as well as the stacked flux (in blue). For groups of objects above the detection limit, \textsc{XID+} has the distinct advantage in having the flux for every objects, compared to stacking which provides an average flux, assuming the underlying flux distribution is Gaussian.

\subsection{Performance with a shallower prior list}
So far, we have shown the performance of \textsc{XID+} with the type of prior lists that are available for the \emph{HerMES} fields (i.e. both relatively deep and selected at a similar wavelength to the SPIRE bands). We now demonstrate the expected performance of \textsc{XID+} for \emph{HATLAS} fields, where the prior list is likely to be derived from shallow, optical ancillary data.

We use the same simulated maps as before, but now our prior list only use sources with an r-band magnitude $<$ 19.8 or a 250 $\mathrm{\mu m}$ flux $>$ 15.48 mJy. With this cut, our prior list contains 5536 sources. Not only does this cut mimic the type of ancillary data available in the \emph{HATLAS} fields (i.e. shallow optical data and additional bright SPIRE sources detected via blind source detection), but is the same cut used to test the Lambda Adaptive Multi-band Deblending Algorithm in R \citep[LAMBDAR][]{Wright:2016}.

\begin{figure*}
\centering 
\subfloat[$250\mathrm{\mu m}$]{\includegraphics[width=5.5cm,page={7}]{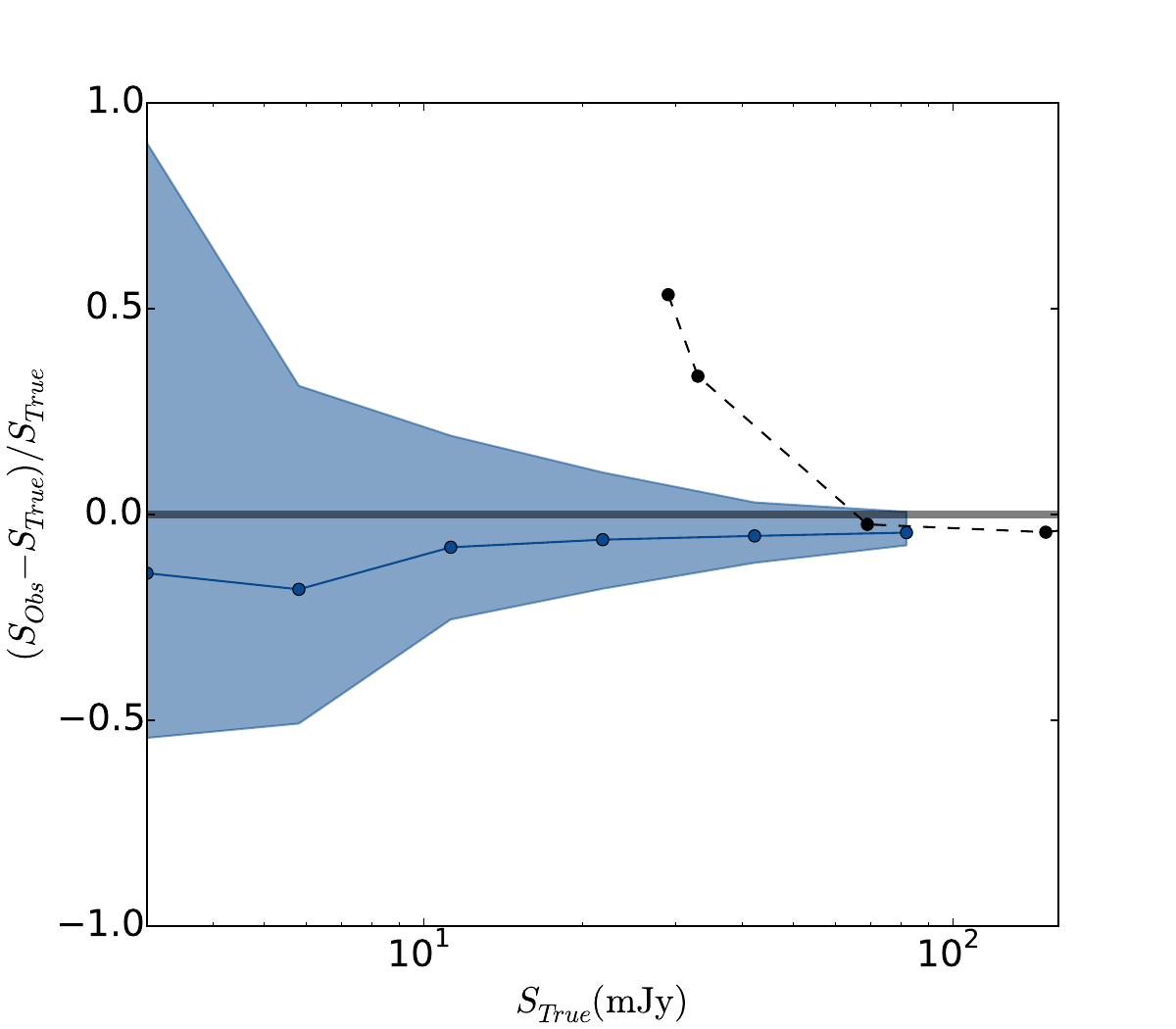}}
\subfloat[$350\mathrm{\mu m}$]{\includegraphics[width=5.5cm,page={8}]{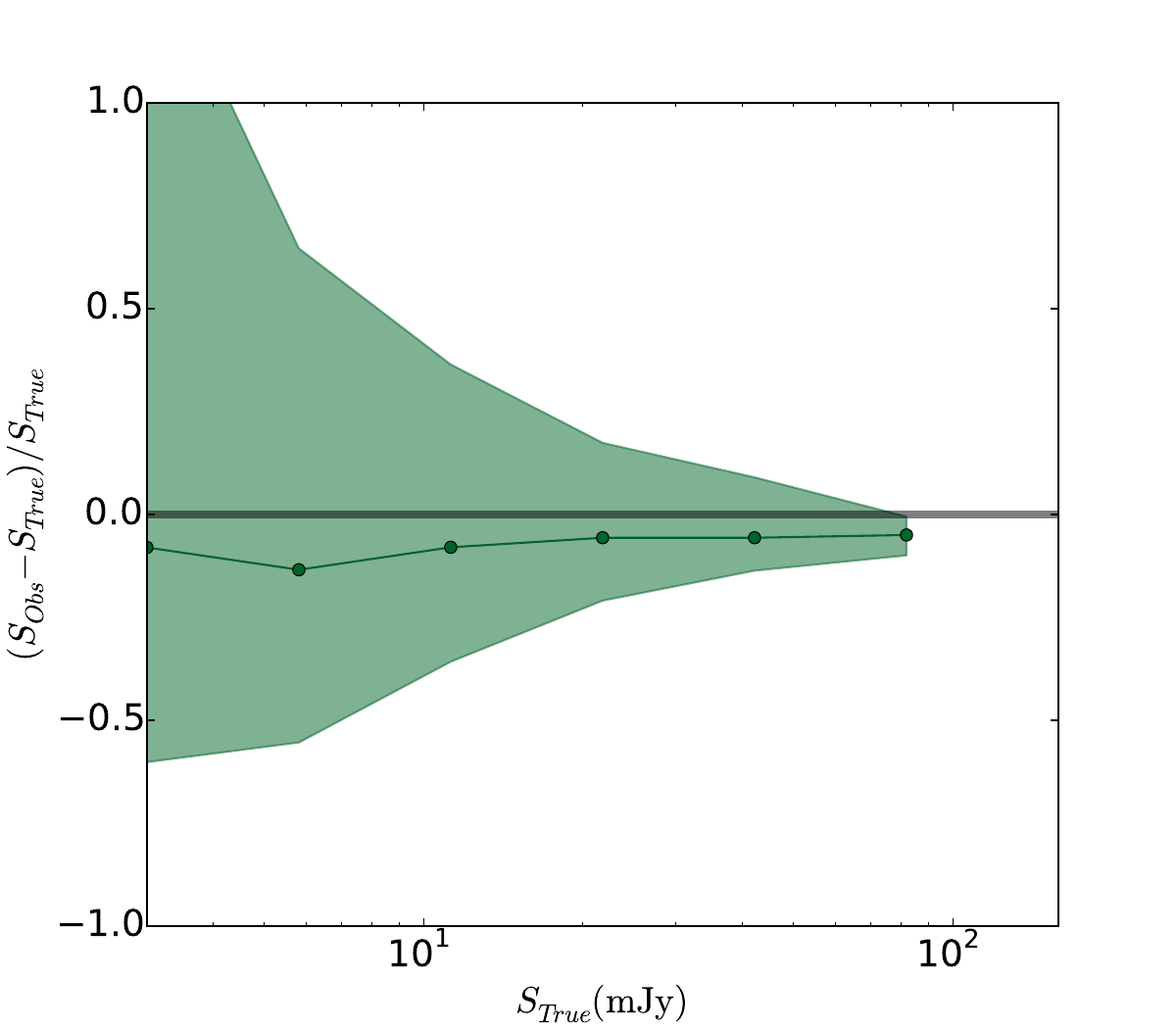}}
\subfloat[$500\mathrm{\mu m}$]{\includegraphics[width=5.5cm,page={9}]{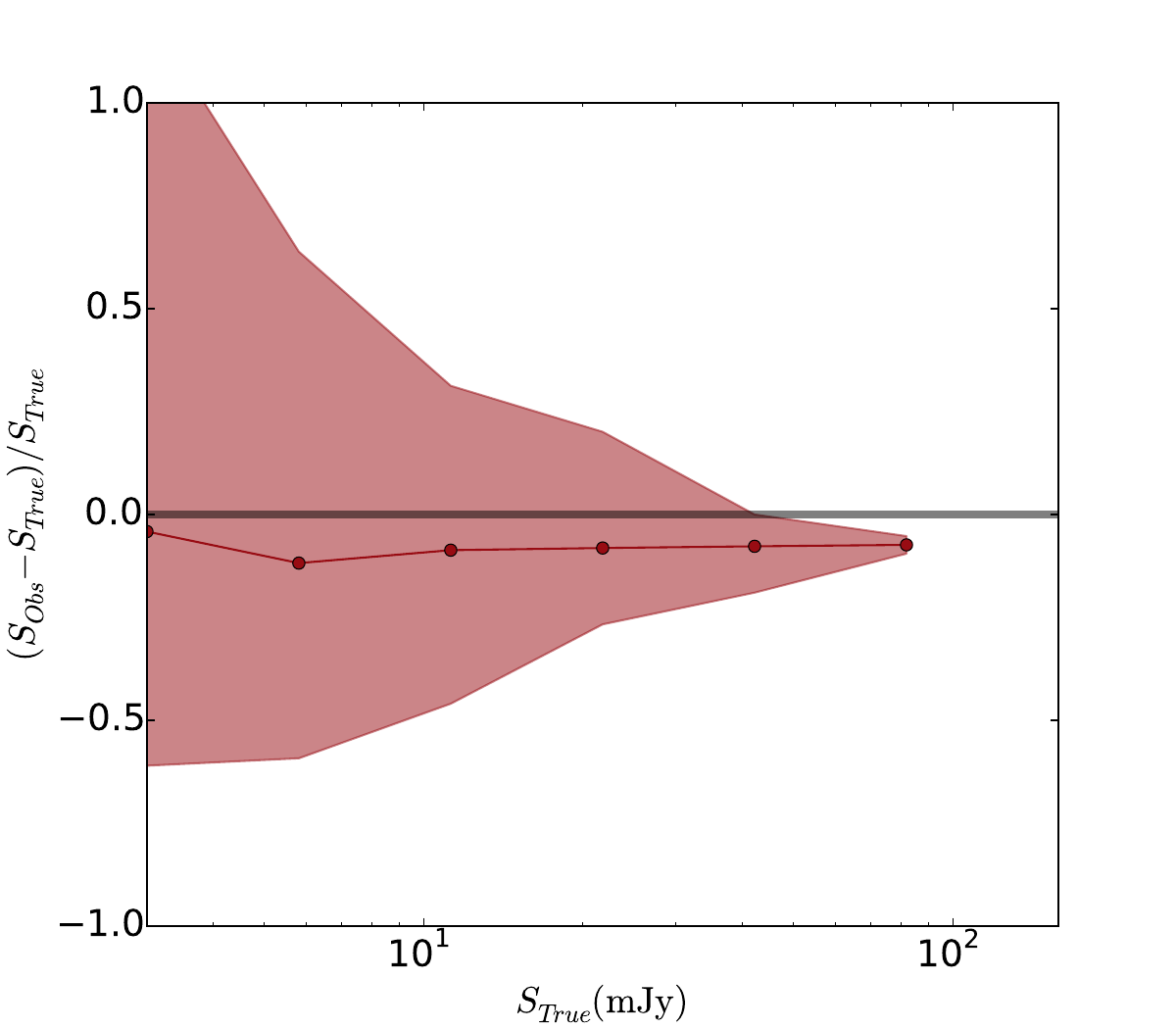}}
\caption{The 16th, 50th and 84th percentiles for flux accuracy of \textsc{XID+} (coloured line and shaded region) and the median flux accuracy for \textsc{LAMBDAR} (black dashed line), as a function of true flux, for the 250 (blue), 350 (green) and 500 (red) $\mathrm{\mu m}$ SPIRE bands. A horizontal thick line is shown at zero for clarity.}\label{fig:accuracy_2}
\end{figure*}

Figure \ref{fig:accuracy_2} shows the flux accuracy of \textsc{XID+}, using the shallower prior list. Although the dispersion in accuracy is worse at the lower fluxes, the median accuracy remains close to zero for all fluxes. For the 250 $\mathrm{\mu m}$, we compare our flux accuracy with that obtained by \textsc{LAMBDAR}. While the flux accuracy of \textsc{XID+} is relatively constant from 10 mJy, the flux accuracy of \textsc{LAMBDAR} only reaches comparable accuracy levels at fluxes greater than 70mJy, illustrating \textsc{XID+} can reach far lower flux levels than \textsc{LAMBDAR}. We note both precision and flux uncertainty accuracy are similar to that obtained in Sections \ref{sec:prec} and \ref{sec:unc}.

\section{COSMOS field}\label{sec:COSMOS}
Having satisfactorily demonstrated the performance on simulations, we have run \textsc{XID+} on the HerMES COSMOS SPIRE maps from the 2nd Data release. As a prior, we take the MIPS 24 $\mathrm{\mu m}$ catalogue \citep{LeFLoch:2009}, which covers an area of 2.265 square degrees and includes 52,092 sources with a 24 $\mathrm{\mu m}$ flux density above 60 $\mathrm{\mu}$Jy, which corresponds to a signal to noise cut of 3.

\begin{figure*}
\centering 
\subfloat[Simulation]{\includegraphics[width=5.5cm]{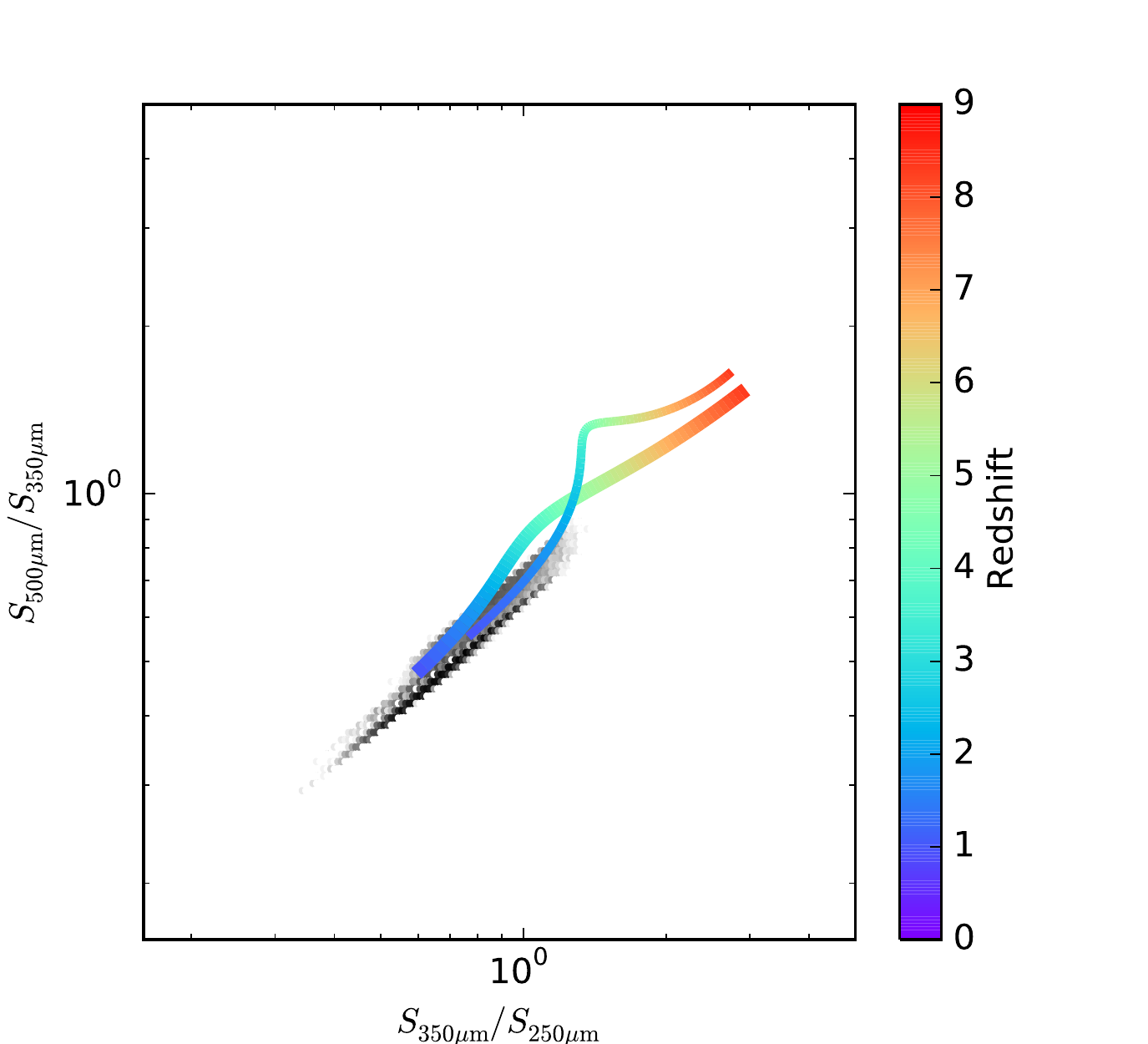}}
\subfloat[\textsc{XID+} fit to simulation]{\includegraphics[width=5.5cm]{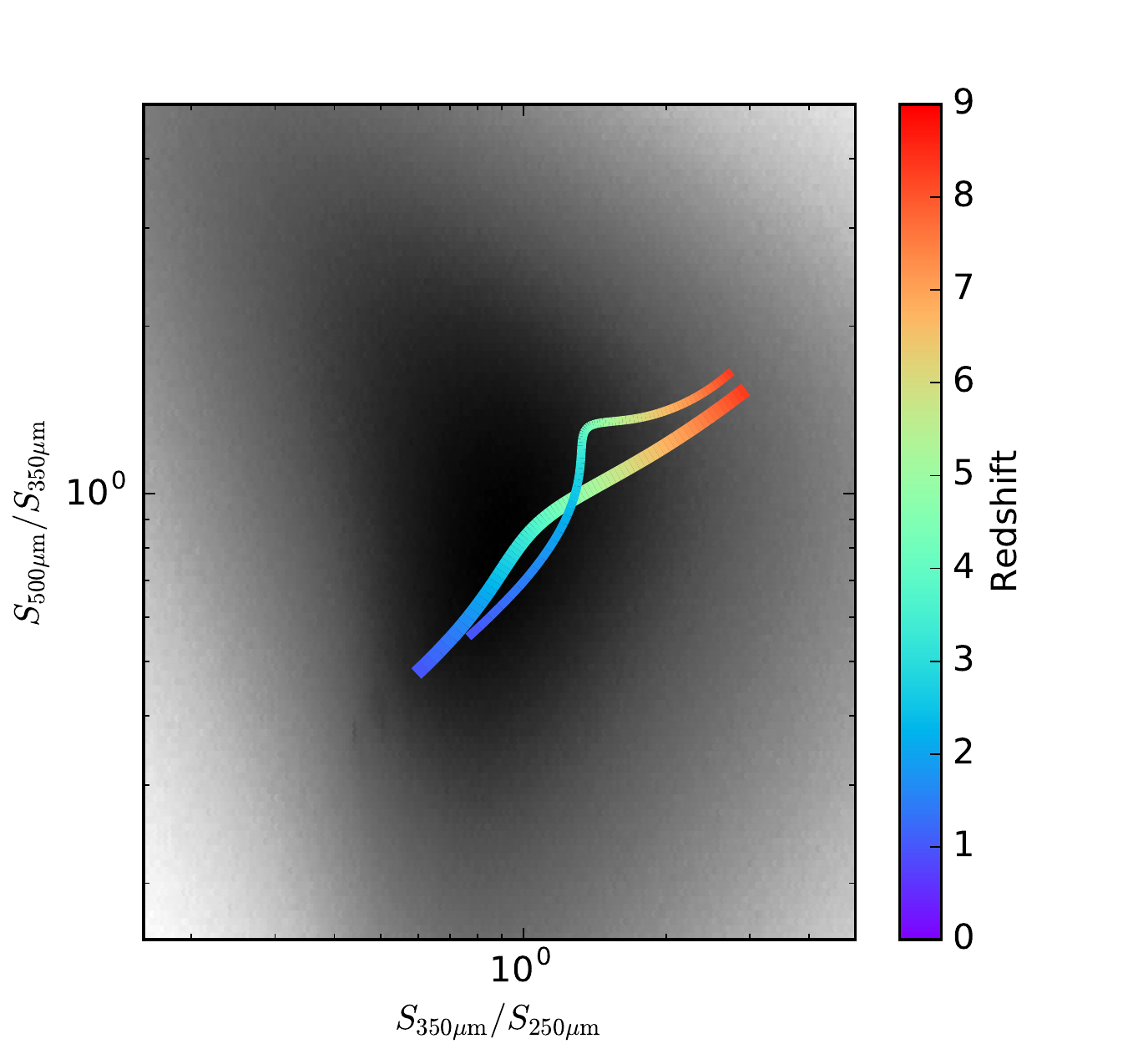}}
\subfloat[\textsc{XID+} fit to COSMOS]{\includegraphics[width=5cm]{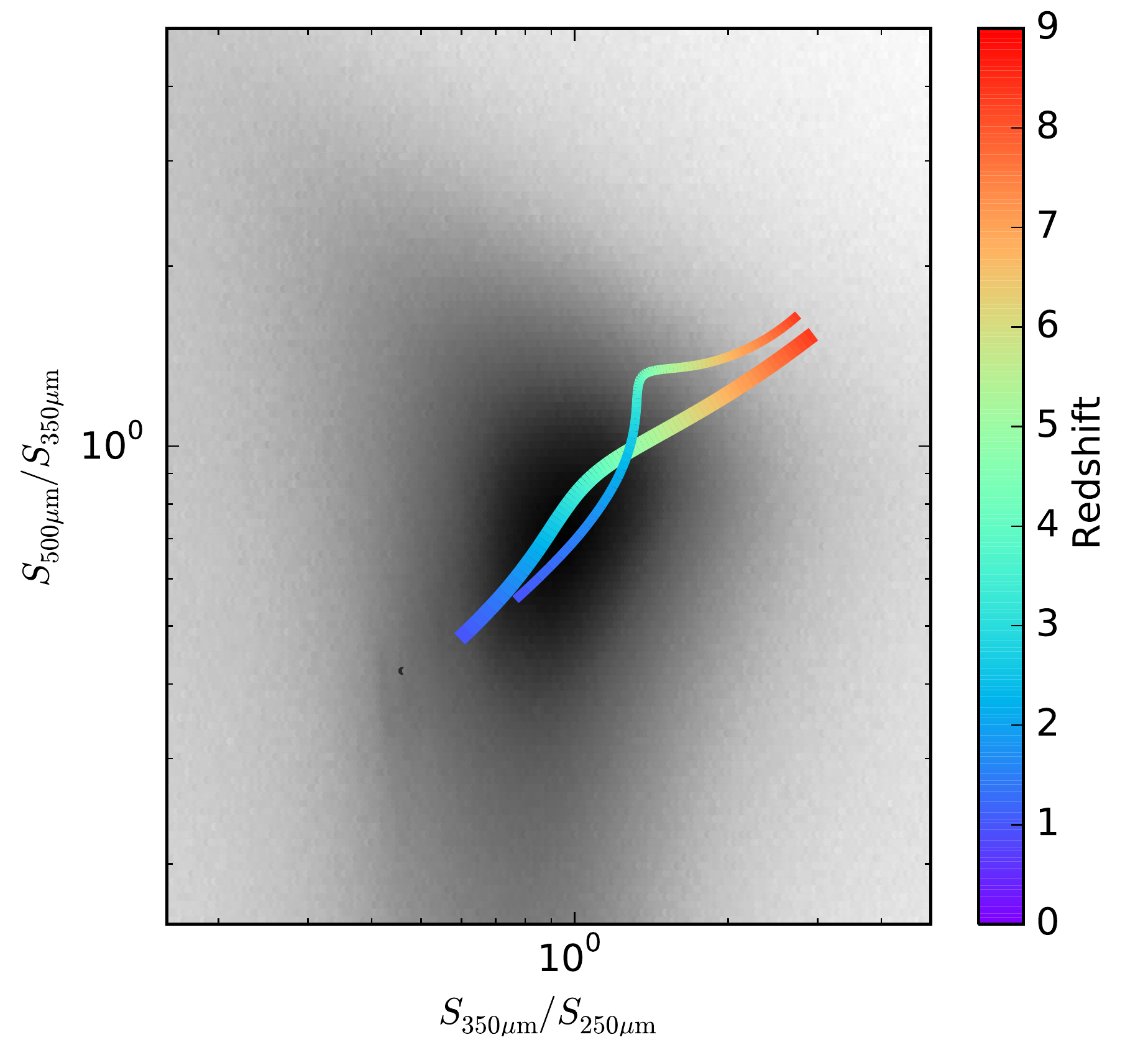}}
\caption{The marginalised SPIRE colour-colour probability density (in black) for the sources used in the simulation (left), \textsc{XID+} fit to simulated sources (centre), and \textsc{XID+} fit of MIPS 24 $\mathrm{\mu m}$ sources in the COSMOS field (right). Over plotted are the redshifted spectral energy distributions for a red star-forming galaxy (thick line) and blue star-forming galaxy (thin line) as defined by \protect\citep{Berta:2013}.}\label{fig:col-col}
\end{figure*}

Figure \ref{fig:col-col} shows the marginalised probability density of our MIPS 24 $\mathrm{\mu m}$ prior catalogue in SPIRE colour-colour space and is  constructed by combining the 1500 samples from the posterior, for all 52,092 sources. The redshift tracks for a red and blue star-forming galaxy spectral energy distribution template, empirically derived from Herschel sources \citep{Berta:2013}, are over-plotted and run through the highest density region at redshifts of around 2-3. As a comparison, we show the true colour distribution for our mock simulation used in section \ref{sec:sims} and the corresponding distribution from the \textsc{XID+} fit. The probability distribution from the fit to the simulation is far wider than the truth. This is not surprising as, at present, the three SPIRE bands are being fit independently, and so any correlation in colour is being dispersed by confusion noise and non-detections. Interestingly, the fit to the real COSMOS data shows a more constrained probability distribution. Whether this is a consequence of slightly different selection, or less variation in SED shape will require further investigation.


\begin{table*}
\begin{tabular}{ | l || c || c || c ||}

\hline
& CIRB $250 \mathrm{\mu m}$ & CIRB $350 \mathrm{\mu m}$ & CIRB $500 \mathrm{\mu m}$ \\
& ($\mathrm{nWm^{-2}sr^{-1}}$) & ($\mathrm{nWm^{-2}sr^{-1}}$) & ($\mathrm{nWm^{-2}sr^{-1}}$) \\
\hline
FIRAS \citep{Lagache:2000}& $11.8\pm2.9$ & $6.4\pm1.6$ & $2.7\pm0.7$ \\
\hline
SPIRE resolved sources \citep{Oliver:2010}& $1.73\pm0.33$ (15\%)& $0.63\pm0.18$ (10\%)& $0.15\pm0.07$ (6\%)\\
\hline
XID+ with $24 \mathrm{\mu m}$ sources& $5.573\pm0.003$ (47\%)& $2.805\pm0.002$ (45\%)& $1.24\pm0.002$ (46\%)\\
\hline
Stacking of $24 \mathrm{\mu m}$ sources \citep{Bethermin:2012}& $7.40^{+1.42}_{-1.43}$ (73\%)& $4.50^{+0.90}_{-0.90}$ (63\%)& $1.54^{+0.30}_{-0.30}$ (55\%)\\
\hline
\end{tabular}
\caption{Contribution to the CIRB at the SPIRE wavelengths from various measures, including the absolute measurement made by FIRAS}\label{tab:cirb}
\end{table*}

Table \ref{tab:cirb} shows the total contribution to the cosmic infrared background (CIRB) at 250, 350 and 500 $\mathrm{\mu m}$ from our MIPS 24 $\mathrm{\mu m}$ prior catalogue, alongside the FIRAS absolute measurements from \cite{Lagache:2000}, the contribution to the CIRB from SPIRE resolved sources \citep{Oliver:2010} and the contribution from stacking MIPS 24 $\mathrm{\mu m}$ sources \citep{Bethermin:2012}. By going to a depth of 60 $\mathrm{\mu}$Jy in the 24 $\mathrm{\mu m}$ catalogue, we reach can explain 47, 44 and 46\% of the nominal measured values at 250, 350 and 500 $\mathrm{\mu m}$ \citep{Lagache:2000}. This compares favourably to the 15, 10 and 6\% that SPIRE resolves at the (40 beams)$^{-1}$ depth \citep{Oliver:2010}. We resolve less of the CIRB than achieved by stacking MIPS 24 $\mathrm{\mu m}$ , with the missing contribution belonging to those sources that \textsc{XID+} cannot constrain. However, unlike stacking, we know how that flux is distributed amongst sources. The remaining  CIRB that is not associated with $\mathrm{\mu m}$ sources, will be coming from other sources not detected at 24 $\mathrm{\mu m}$.

Our final data product consists of a catalogue, summarising the SPIRE fluxes via the 16th, 50th and 84th percentile (i.e. median and median $\pm \sigma$), the median background and the convergence statistics. We also make available on request, the 3000 samples from the posterior probability distribution, each of which can be thought of as a possible catalogue in probability space.

\section{Discussion}\label{sec:disc}
We have run \textsc{XID+} on simulated maps of the COSMOS field and compared it to \textsc{DESPHOT} using three main metrics; flux accuracy, precision and uncertainty accuracy. On accuracy, \textsc{XID+} performs significantly better in all three bands, and although appearing  marginally less precise, the loss of precision relates to more realistic estimates for flux uncertainties. 

The higher performance gained by \textsc{XID+} comes from fully capturing the posterior probability distribution on flux estimates. By exploring the posterior, we get a proper handle on uncertainties and no longer have to employ penalisation techniques such as LASSO, which are known to behave erratically.

By using a probabilistic approach, we have a framework where we can introduce prior information on the source fluxes in a transparent manner. For this basic version, we use a simple $\log_{10}$ uniform prior, with bounds at $10^{-2}$ and $10^{3}\mathrm{mJy}$. However, as can be seen in Figure \ref{fig:corr}, where sources are correlated, if we have prior knowledge on the flux of one of the sources, it can help us determine a more precise flux for the other.

As demonstrated by \cite{Safarzadeh:2015}, these priors could come from fitting spectral energy distribution models to multi-wavelength ancillary data. Another alternative is to use machine-learning algorithms to `learn' the expected flux from the statistical population. As part of HELP, the testing and benchmarking of suitable methods for deriving SPIRE flux priors will be presented in Hurley et al. (in prep). 

More generally, the probabilistic model used in \textsc{XID+} can easily be expanded, allowing distributions such as the flux distribution (or number counts) to be modelled explicitly. In principle, and with additional information such as redshifts, the probabilistic model could become detailed enough to simultaneously fit luminosity functions and the location of locus of the SFR-M* relation at different redshift and we will explore these expansions in future papers.

\section{Conclusions}\label{sec:conc}
In this paper we have introduced the prior based source detection software, \textsc{XID+}. By using the Bayesian inference tool \emph{Stan}, we are able to fully sample the posterior probability distribution, which in turn gives a better understanding of the uncertainty associated with the source flux. 

Having run \textsc{XID+} on simulated maps, we have shown this is extremely advantageous for maps that are confusion limited, such as the \emph{Herschel} observations that are part of \emph{HerMES}. In comparison to the current maximum likelihood based software \textsc{DESPHOT}, XID+ performs far better in all three main metrics; flux accuracy, precision and uncertainty accuracy.

We have run \textsc{XID+} on the HerMES COSMOS SPIRE maps from the 2nd Data release, using the MIPS 24 $\mathrm{\mu m}$ catalogue \citep{LeFLoch:2009} as our prior. Using the full posterior, we have created a marginalised SPIRE colour-colour plot, illustrating the probability distribution of our MIPS 24 $\mathrm{\mu m}$ catalogue in SPIRE colour-colour space. We have also shown that the MIPS 24 $\mathrm{\mu m}$ sources contribute 47, 44 and 46\%  to the cosmic infrared background at 250, 350 and 500 $\mathrm{\mu m}$. We provide the catalogue and posterior probability distribution samples as a data product as part of HELP \footnote{Data products will be made publicly available at the end of HELP in December 2017, however early access can be granted on request.}. As far as we are aware, this is the first time the full posterior probability distribution is made available as a data product for list driven photometry.

\section*{Acknowledgements}%
The research leading to these results has received funding from the European Union Seventh Framework Programme FP7/2007-2013/ under grant agreement no.607254. This publication reflects only the author's view and the European Union is not responsible for any use that may be made of the information contained therein. 

Seb Oliver acknowledges support from the Science and Technology Facilities Council (grant numberST/L000652/1) and Michael Betancourt is supported under EPSRC grant EP/J016934/1.

HCSS/HSpot/HIPE are joint developments by the Herschel Science Ground Segment Consortium, consisting of ESA, the NASA Herschel Science Center, and the HIFI,PACS and SPIRE consortia.

SPIRE has been developed by a consortium of institutes led by Cardiff Univ. (UK) and including Univ. Lethbridge(Canada); NAOC (China); CEA, LAM (France); IFSI, Univ.Padua (Italy); IAC (Spain); Stockholm Observatory (Sweden); Imperial College London, RAL, UCL-MSSL, UKATC,Univ.  Sussex  (UK);  Caltech, JPL, NHSC, Univ.  Colorado(USA).  This development has been supported by national funding agencies: CSA (Canada); NAOC (China);  CEA,CNES, CNRS (France); ASI (Italy); MCINN (Spain); SNSB(Sweden); STFC, UKSA (UK); and NASA (USA).
Special thanks goes to Yannick Roehlly for organisation of code, and Louise Winters for management within HELP. This paper make use of \textsc{Matplotlib} \citep{Hunter:2007} and \textsc{Astropy}, a community-developed core Python package for Astronomy \citep{Astropy-Collaboration:2013}. 

\bibliography{HELP_bib}
\appendix
\renewcommand{\thefigure}{A\arabic{figure}}

\setcounter{figure}{0}
\section*{Appendix A}\label{Stan_model}
\onecolumn
Stan code for XID+.
\begin{lstlisting}[language=C, linewidth=18cm]
//Full Bayesian inference fit  XID
data {
  int<lower=0> nsrc;//number of sources
  //----PSW----
  int<lower=0> npix_psw;//number of pixels
  int<lower=0> nnz_psw; //number of non neg entries in A
  vector[npix_psw] db_psw;//flattened map
  vector[npix_psw] sigma_psw;//flattened uncertianty map (assuming no covariance between pixels)
  real bkg_prior_psw;//prior estimate of background
  real bkg_prior_sig_psw;//sigma of prior estimate of background
  vector[nnz_psw] Val_psw;//non neg values in image matrix
  int Row_psw[nnz_psw];//Rows of non neg valies in image matrix
  int Col_psw[nnz_psw];//Cols of non neg values in image matrix
  vector[nsrc] f_low_lim_psw;//upper limit of flux 
  vector[nsrc] f_up_lim_psw;//upper limit of flux 
  //----PMW----
  int<lower=0> npix_pmw;//number of pixels
  int<lower=0> nnz_pmw; //number of non neg entries in A
  vector[npix_pmw] db_pmw;//flattened map
  vector[npix_pmw] sigma_pmw;//flattened uncertianty map (assuming no covariance between pixels)
  real bkg_prior_pmw;//prior estimate of background
  real bkg_prior_sig_pmw;//sigma of prior estimate of background
  vector[nnz_pmw] Val_pmw;//non neg values in image matrix
  int Row_pmw[nnz_pmw];//Rows of non neg valies in image matrix
  int Col_pmw[nnz_pmw];//Cols of non neg values in image matrix
  vector[nsrc] f_low_lim_pmw;//upper limit of flux (in log10)
  vector[nsrc] f_up_lim_pmw;//upper limit of flux (in log10)
  //----PLW----
  int<lower=0> npix_plw;//number of pixels
  int<lower=0> nnz_plw; //number of non neg entries in A
  vector[npix_plw] db_plw;//flattened map
  vector[npix_plw] sigma_plw;//flattened uncertianty map (assuming no covariance between pixels)
  real bkg_prior_plw;//prior estimate of background
  real bkg_prior_sig_plw;//sigma of prior estimate of background
  vector[nnz_plw] Val_plw;//non neg values in image matrix
  int Row_plw[nnz_plw];//Rows of non neg valies in image matrix
  int Col_plw[nnz_plw];//Cols of non neg values in image matrix
  vector[nsrc] f_low_lim_plw;//upper limit of flux 
  vector[nsrc] f_up_lim_plw;//upper limit of flux 

}


parameters {
  vector<lower=0.0,upper=1.0>[nsrc] src_f_psw;//source vector
  real bkg_psw;//background
  vector<lower=0.0,upper=1.0>[nsrc] src_f_pmw;//source vector
  real bkg_pmw;//background
  vector<lower=0.0,upper=1.0>[nsrc] src_f_plw;//source vector
  real bkg_plw;//background
  real<lower=0.0,upper=8> sigma_conf_psw;
  real<lower=0.0,upper=8> sigma_conf_pmw;
  real<lower=0.0,upper=8> sigma_conf_plw;

}

model {
  vector[npix_psw] db_hat_psw;//model of map
  vector[npix_pmw] db_hat_pmw;//model of map
  vector[npix_plw] db_hat_plw;//model of map


  vector[npix_psw] sigma_tot_psw;
  vector[npix_pmw] sigma_tot_pmw;
  vector[npix_plw] sigma_tot_plw;

  vector[nsrc] f_vec_psw;//vector of source fluxes
  vector[nsrc] f_vec_pmw;//vector of source fluxes
  vector[nsrc] f_vec_plw;//vector of source fluxes
  // Transform to normal space. As I am sampling variable then transforming I don't need a Jacobian adjustment
  for (n in 1:nsrc) {
    f_vec_psw[n] <- f_low_lim_psw[n]+(f_up_lim_psw[n]-f_low_lim_psw[n])*src_f_psw[n];
    f_vec_pmw[n] <- f_low_lim_pmw[n]+(f_up_lim_pmw[n]-f_low_lim_pmw[n])*src_f_pmw[n];
    f_vec_plw[n] <- f_low_lim_plw[n]+(f_up_lim_plw[n]-f_low_lim_plw[n])*src_f_plw[n];


  }
  //Prior on background 
  bkg_psw ~normal(bkg_prior_psw,bkg_prior_sig_psw);
  bkg_pmw ~normal(bkg_prior_pmw,bkg_prior_sig_pmw);
  bkg_plw ~normal(bkg_prior_plw,bkg_prior_sig_plw);
 
   
  // Create model maps (i.e. db_hat = A*f) using sparse multiplication
  for (k in 1:npix_psw) {
    db_hat_psw[k] <- bkg_psw;
    sigma_tot_psw[k]<-sqrt(square(sigma_psw[k])+square(sigma_conf_psw));
  }
  for (k in 1:nnz_psw) {
    db_hat_psw[Row_psw[k]+1] <- db_hat_psw[Row_psw[k]+1] + Val_psw[k]*f_vec_psw[Col_psw[k]+1];
      }

  for (k in 1:npix_pmw) {
    db_hat_pmw[k] <- bkg_pmw;
    sigma_tot_pmw[k]<-sqrt(square(sigma_pmw[k])+square(sigma_conf_pmw));
  }
  for (k in 1:nnz_pmw) {
    db_hat_pmw[Row_pmw[k]+1] <- db_hat_pmw[Row_pmw[k]+1] + Val_pmw[k]*f_vec_pmw[Col_pmw[k]+1];
      }

  for (k in 1:npix_plw) {
    db_hat_plw[k] <- bkg_plw;
    sigma_tot_plw[k]<-sqrt(square(sigma_plw[k])+square(sigma_conf_plw));
  }
  for (k in 1:nnz_plw) {
    db_hat_plw[Row_plw[k]+1] <- db_hat_plw[Row_plw[k]+1] + Val_plw[k]*f_vec_plw[Col_plw[k]+1];
      }

  // likelihood of observed map|model map
  db_psw ~ normal(db_hat_psw,sigma_tot_psw);
  db_pmw ~ normal(db_hat_pmw,sigma_tot_pmw);
  db_plw ~ normal(db_hat_plw,sigma_tot_plw);

    }
\end{lstlisting}
\section*{Appendix B}\label{stacking_app}
\clearpage
\begin{figure}
\onecolumn
\includegraphics[width=18cm]{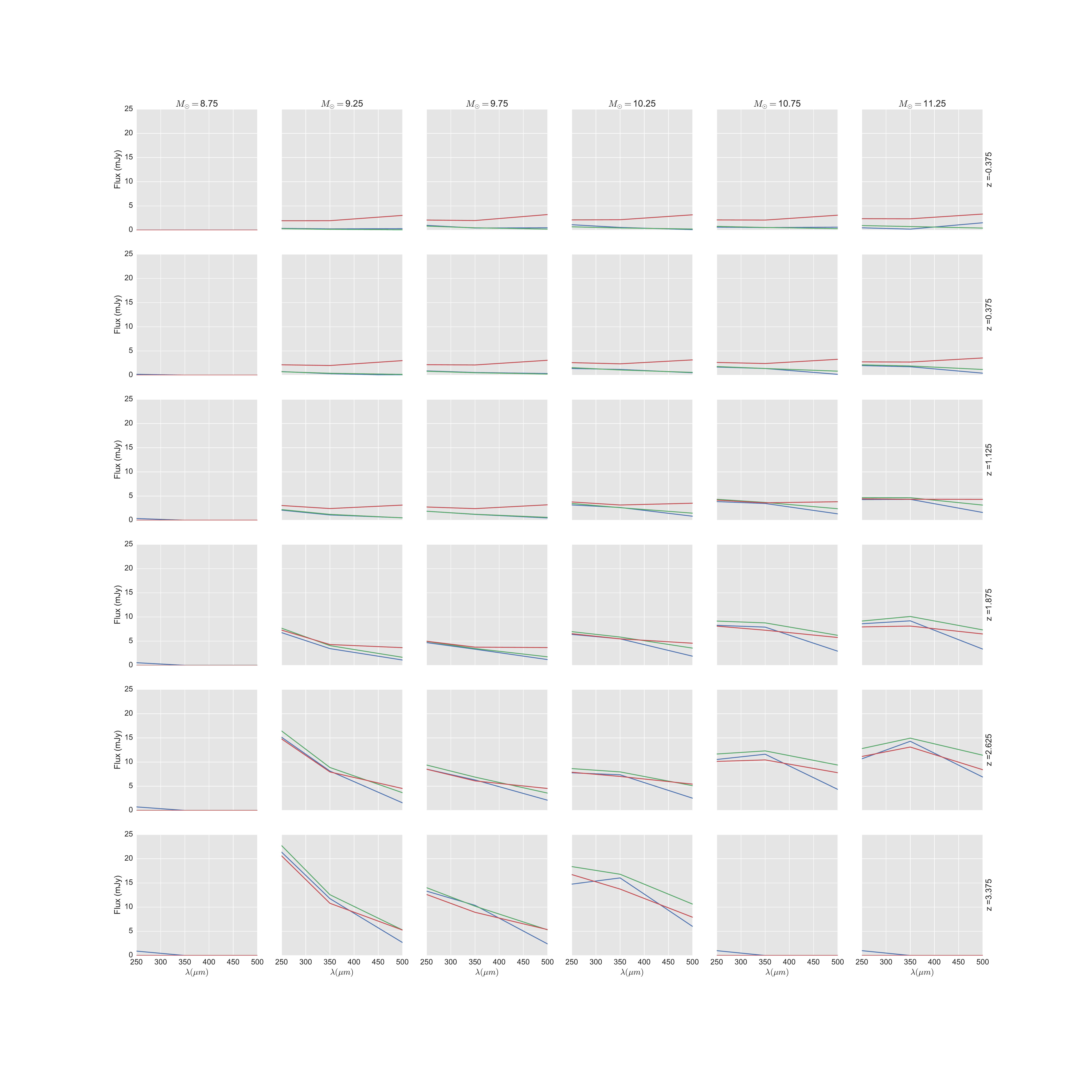}
\caption{The average flux from simulated objects, grouped by Stellar Mass and redshift. The flux from stacking is shown in blue, the average true flux for each group in green, and the average \textsc{XID+} flux for the group in red. For average fluxes above the detection limit, \textsc{XID+} is just as capable of returning true average flux as stacking, but with the added advantage of having the flux for each object.}\label{fig:stack}
\end{figure}

%
%
%
%
%
%

\end{document}